\documentclass[letterpaper,11pt]{article}
\usepackage[letterpaper, left=1in, top=1in, right=1in, bottom=1in,nohead,includefoot]{geometry}
\usepackage{color,charter,graphicx,natbib,enumitem,amsmath,amssymb,amsfonts,titlesec,placeins,bm}
\usepackage{sidecap} \sidecaptionvpos{figure}{c} 
\usepackage[svgnames,x11names]{xcolor}
\usepackage[pdftex]{hyperref}
	\hypersetup{
	colorlinks,%
	linkcolor=MidnightBlue,  
	urlcolor=DarkSlateGray,   
	citecolor=RoyalBlue2  
	}
\titleformat*{\section}{\normalfont\Large\bfseries\blu}
\titleformat*{\subsection}{\normalfont\large\bfseries\blu}
\titleformat*{\subsubsection}{\normalfont\normalsize\bfseries\blu}
\def\parablu#1{\paragraph{\blu #1}}
 
\def\blu{\color{RoyalBlue4}}      
\def\bone{\mathbf{1}}\def\bzero{\mathbf{0}}

\def\I{\mathbf{I}}\def\S{\mathbf{S}}\def\V{\mathbf{V}}
\def\c{\mathbf{c}}\def\e{\textrm{e}}\def\f{\mathbf{f}}\def\p{\mathbf{p}}\def\s{\mathbf{s}}
\def\y{\mathbf{y}}

\def\btau{{\bm\tau}}\def\bepsilon{{\bm\epsilon}}\def\btheta{{\bm\theta}} \def\bphi{{\bm\phi}} 
\def\bLambda{{\bm\Lambda}}

\def\seq#1#2{#1{:}#2}\def\j1J{j=\seq 1J}
\def\eqn#1{eqn.~(\ref{eq:#1})}
\def\diag{\textrm{diag}}

\def\bi{\begin{itemize}[noitemsep,topsep=3pt]}
\def\ei{\end{itemize}}\def\i{\item}
\def\bn{\begin{enumerate}[noitemsep,topsep=3pt]}
\def\en{\end{enumerate}}
\def\beq#1{\begin{equation}\label{eq:#1}}\def\eeq{\end{equation}}
\def\beas{\begin{eqnarray*}}\def\eeas{\end{eqnarray*}}
\def\bea{\begin{eqnarray}}\def\eea{\end{eqnarray}}

\def\figdir{./}    

\begin{document}
 
\title{\blu On Entropic Tilting and Predictive Conditioning} 
\author{Emily Tallman and Mike West\\ Duke University \\ 
\small \href{emily.tallman@duke.edu}{emily.tallman@duke.edu}, \href{mike.west@duke.edu}{mike.west@duke.edu}}
\def\today{\small {\em Original version:} June 1, 2022 \\ {\em Current version:}  August 14, 2022}
\maketitle
\begin{abstract}
Entropic tilting (ET) is a Bayesian decision-analytic method for constraining distributions to satisfy defined targets or bounds for sets of expectations. This report recapitulates the foundations and basic theory of ET for conditioning predictive distributions on such constraints, recognising   the increasing interest in ET  in several application areas.  Contributions  include new results related to connections with regular exponential families of distributions, and  the extension of ET to {\em relaxed entropic tilting (RET)} where specified  values for expectations define bounds rather than exact targets. Additional new developments include theory and examples that condition on quantile constraints for modified predictive distributions and examples relevant to Bayesian forecasting applications. 
\end{abstract}
 
\section{ET Background, Theory and Insights}

ET is Bayesian decision-analytic and variational method for conditioning predictive distributions on defined sets of constraints, and is becoming of increasing interest  in econometrics, finance and business applications~\citep[e.g.][]{RobertsonET2005,KrugerET2017,West2021decisionconstraints,TallmanWest2022} among other areas.  The basic definitions, properties, and some new theoretical results and insights are noted here.

\subsection{Setting: ET Constraints and Decision}

Suppose $\y$ is a random $m-$vector with (predictive) p.d.f. $p(\y)$. The random vector may be discrete, continuous, or mixed.  Notation for expectations will ignore mathematical details and play-out as if $\y$ is continuous with respect to Lebesgue measure, for simplicity and clarity of presentation. 
Refer to $p(\y)$ as the p.d.f. of the {\em baseline distribution} for $\y$.  

 Now consider all possible p.d.f.s $f(\y)$ with the same support as $p(\y)$. A set of {\em moment constraints} is to be assessed and/or imposed.  In the standard formulation of ET, these specify exact \lq\lq target'' values of a set of expectations under any p.d.f. $f(\y)$ that is consistent with these values.  That is, define  the expectations 
$E_f(s_j(\y)) = s_j$, for $j=\seq 1q$ for some chosen set of $q$  {\em score} functions  $s_j(\y)$  with specified target/constrained values $s_j$.  Here the expectations are with respect to any chosen  p.d.f. $f(\y)$ that satisfies these constraints.        It is convenient for theoretical development to take $s_1(\y)=1$ and $s_1=1$ to represent the fact that $f(\y)$ is a p.d.f.,  i.e., to define the required normalization constraint, although this is not necessary in implementations.   The other constraints are problem specific.   
In vector notation, the constraints are $E_f(\s(\y)) = \s $ with $q-$vectors $\s(\y) = (s_1(\y),\ldots,s_q(\y))'$ and  target vector $\s = (s_1,\ldots,s_q)'.$ 

The standard ET framework defines the following decision problem: 
choose $f(\y)$ as the K\"ullback-Leibler closest  p.d.f. to the baseline $p(\y)$ subject to these constraints. 
ET uses the KL divergence \underbar{\blu of} $p(\y)$ \underbar{\blu from} $f(\y)$,    $$ {KL}_{p|f} = \int_\y \log(f(\y)/p(\y))f(\y)d\y.$$   The connotation is that $f(\y)$ is best/optimal/desired and $p(\y)$ is a step towards it/an approximation, since $f(\y)$ is the anchor distribution in this KL direction.  This is the same \lq\lq direction'' as used in variational Bayes' methods. The \lq\lq other direction'' is more natural and relevant in other settings~\citep[e.g. such as used in sequential Bayesian analysis and forecasting in simultaneous graphical dynamic linear models in][]{GruberWest2016BA,GruberWest2017ECOSTA}.

\subsection{ET Solution} 

Assuming that a solution exists, it has the unique form  
\begin{equation}\label{eq:fETpdf} 
f(\y) \propto p(\y) \exp(\btau'\s(\y))  =   k_{\btau} p(\y) \exp(\btau'\s(\y)) 
\end{equation}
where  $\btau'\s(\y) = \sum_{j=1{:}q} \tau_j s_j(\y),$   the vector
 $\btau = (\tau_1,\ldots,\tau_q)'$ is a parameter that ensures $f(\y)$ satisfies the specified expectation constraints, and 
$k_{\btau}$ is the normalizing constant.  

Existence of the solution depends on the form of $p(\y)$ and of the score functions $s_j(\y)$ (and, in some cases, on the choices of values of target/expected scores $s_j$ for given choices of score functions).  It is easy to see that some choices of score functions can lead to no feasible solution.  As easy examples, if $p(\y)$ is Cauchy or log-T and $\s(\y)=\y,$ then no solution exists.  Other cases are more subtle: existence of a solution corresponds to the existence of  the moment generating function of $\s(\y)$ when $\y\sim p(\y).$  It is obvious that a solution exists when the score functions are bounded in $\y,$ and the role of scores as utility functions encourages choices of bounded forms as natural in some applications.  

Whatever the forms of the scores,  the following discussion assumes that a solution exists.   For a given $\btau,$ the ET p.d.f. of 
 \eqn{fETpdf} can be written in the exponential family form 
\begin{equation}\label{eq:fETEXPF} 
f(\y) =  p(\y) \exp\{\btau'\s(\y) - c(\btau)\}\quad\textrm{where}\quad   c(\btau) = -\log( k_{\btau}).  
\end{equation}
It trivially follows that 
\begin{equation}\label{eq:KLpfs}   {KL}_{p|f} \equiv \kappa(\s) = \btau'\s - c(\btau)\end{equation}
where $\s  = E_f(\s(\y)).$    Hence, in the ET analysis    
at the  specified target $\s,$   the implied optimal $\btau\equiv\btau(\s)$ delivers the minimised value of the KL divergence as $\kappa(\s).$ 

\newpage

\subsection{Comments and Perspectives on ET Settings} 

The analysis requires specification of score functions $\s(\y)$ and target values $\s.$ For the former, 
It is apparent that ET can be used to constrain/adapt a  p.d.f. to have a given mean and/or variance, or to have a set of specified quantiles using indicator score functions.  Examples are given in  Sections~\ref{sec:TExamples} and ~\ref{sec:QExamples} below. 

ET analysis has major roles in exploring/interrogating whether sets of specified constraints are consistent with the baseline, and in exploring \lq\lq perturbations'' of the baseline through choices of target scores $\s$ that represent modest deviations from that implied by the baseline.  Define 
 $\s_{\bzero} = E_p(\s(\y))$  to be the expected score under the baseline, noting that this corresponds to $\btau=\bzero$,     $c(\bzero) = 0$ and, of course,  
$\kappa(\bzero)=0$ is the absolute minimum KL divergence.     Then, assuming $\s(\y)$ is a vector of utilities such that higher values are preferred, then choosing a target score vector $\s = \s_{\bzero}+\bepsilon$ for some vector $\bepsilon$ with positive elements tilts the baseline towards a p.d.f. $f(\y)$ with larger expected utilities. Taking \lq\lq small'' positive entries in $\bepsilon,$-- perhaps anchored as percent increases of values in $\s_{\bzero}$-- defines a perspective of exploring small perturbations of the baseline.   
 
\subsection{Bayes' Factor Interpretation} 
The above development provides entr\'ee to mapping the KL divergence to an interpretable scale.   Suppose $f(\y)$ represents the known distribution of $\y$ and $p(\y)$ a different distribution, linked via the ET construction.   Then, explicitly, an observation at a value $\y$ leads to the implied log 
Bayes' factor in favour of $f(\cdot)$ over $p(\cdot)$ given by $\btau'\s(\y)-c(\btau).$   Hence the minimum KL value 
$\kappa(\s)$ is   precisely the {\em expected log Bayes' factor} in favour of $f(\cdot)$ from a future $\y\sim f(\cdot).$    This anchors interpretation to values and ranges of log Bayes' factors (a.k.a. log likelihood ratios) in this ``two hypotheses'' setting, by mapping to the absolute probability scale; e.g., converting to the implied posterior probability on $f(\cdot)$ under a 50:50 prior. The realised $\kappa(\s)$ in any ET analysis can then be calibrated to the probability scale for communication.

\subsection{Implementing ET} 

Computing the value of 
$\btau = (\tau_1,\ldots,\tau_q)'$  to normalize and satisfy the (other) expectation constraints delivers the result in any given example. 
At the ET optimal $f(\y) \propto p(\y) \exp(\btau'\s(\y)),$  the constraints are 
$$ \s = E_f(\s(\y)) = \int_\y \s(\y) f(\y)d\y$$
in the general notation.  This is equivalent to $\bphi(\btau)=\bzero$ where  the $q-$vector function $\bphi(\cdot)$  is
\bea \label{eq:ETtau} \bphi(\btau) =  \int_\y \{ \s(\y) -\s \} \exp(\btau'\s(\y)) p(\y)  d\y. \eea
There are many simple examples as well as practically important examples where the optimizing value of $\btau$ can be evaluated analytically. In other settings,   finding the optimal $\btau$ vector can often be done using Newton-Raphson or other standard algorithms to solve $\bphi(\btau)=\bzero.$   
Simulation from $p(\y)$ will often be used to provide Monte Carlo approximations to the required expectations here.

\subsection{Simulation of ET Distribution} 

As noted above, simulation from the baseline distriubution will often be used to provide Monte Carlo approximations to expectations in the ET analysis. Given an ET solution,  this naturally leads to importance sampling (IS) as a main tool in summarising predictions under the ET p.d.f. $f(\y).$ This was pointed out by~\cite{RobertsonET2005}; see more recent examples in \cite{West2021decisionconstraints} and \cite{TallmanWest2022}. 

Suppose a Monte Carlo (MC) random sample ${\cal Y}_{p,I} = \{ \y_i, \ i=\seq 1I \}$ of size $I.$ is drawn from the 
baseline distribution. 
Expectations under $p(\y)$ are approximated by sample averages over the MC sample ${\cal Y}_{p,I}$. 
The ET analysis defines the modified predictive distribution $f(\y)$ in which the MC sample is reweighted: each sample $\y_i$ has IS weight $w_i \propto \exp\{\btau'\s(\y_i)\} $  and $\sum_{i=\seq 1I} w_i =1.$  
Standard IS ideas and methodology then apply.  For example, the effective sample size $ESS = 1/\sum_{i=\seq 1I}w_i^2$ is one measure of how \lq\lq close''  $p(\y)$ is to $f(\y),$ representing one numerical measure of how consistent $p(\y)$ is with the imposed ET constraints. Further, the weighted IS sample can be converted to an equally-weighted sample from $f(\y)$ (usually with repeat values, of course) simply by resampling (with replacement) from the set ${\cal Y}_{p,I}$ according to the weights $w_i.$ 
Then expectations under $f(\y)$ are then approximated by simple averages over the resampled set.

\subsection{Relaxed Entropic Tilting (RET)}  

An important contribution of this report is the result that the ET result is relevant-- and optimal-- under broader considerations.  Specifically, the results from the modified perspective in which  $\s$  define {\em bounded expectations} on score functions, not just explicit/exact expectation constraints.    See this as follows. 

Again suppose (with no loss of generality) that each score vector element $s_j(\y)$  represents a utility dimension such that higher values are preferable.  Now consider relaxed moment constraints
 $E_f(\s(\y)) \ge \s > \s_{\bzero}$, i.e.,  elementwise, $E_f(s_j(\y))\ge s_j > s_{0,j}$, for a specified {\em lower bound} target vector $\s$. Here   
 $\s_{\bzero} = E_p(\s(\y))$  corresponds to $\btau=\bzero$,     $c(\bzero) = 0$ and, of course,  
$\kappa(\bzero)=0$ is the absolute minimum KL divergence.   

\begin{SCfigure}[1.15][htp!] 
\centering
\hskip-.15in\includegraphics[width=2.9in]{\figdir/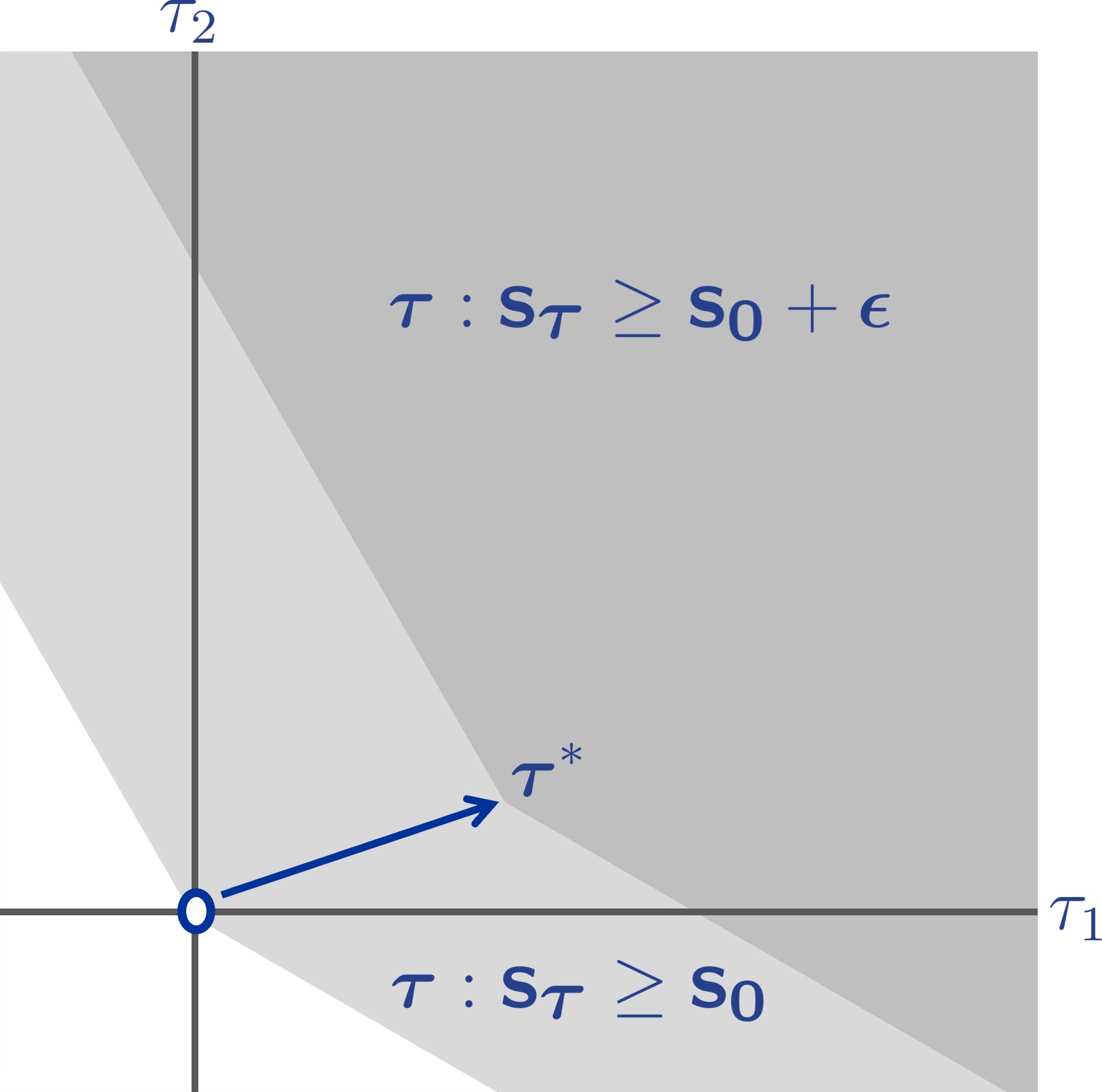} 
    \caption{RET schematic.  The graph represents the space of ET distributions $f(\y)$ indexed by tilting vector $\btau,$ with $\btau=\bzero$ at the baseline $p(\y).$   The union of the shaded regions identifies $\btau$ such that the implied expected scores $\s_{\btau} = E_f(\s(\y)) $ exceed $\s_{\bzero}$ elementwise.  The darker shaded subregion refines this, identifying $\btau$  such that
$\s_{\btau}\ge  \s_{\bzero}+\bepsilon$ for some non-negative vector $\bepsilon$ with at least one positive element. The point $\btau^*$ represents the unique distribution with expected score exactly $\s_{\btau^*} = \s_{\bzero}+\bepsilon$; this is the KL$-$closest to the baseline among all ET distributions with expected scores $ \s_{\btau}\ge \s_{\bzero}+\bepsilon.$  The piecewise linear boundaries here represent the specific bivariate normal, linear score example of Section~\ref{sec:egbivarnormalmean}; in other examples the boundaries are convex and maybe piecewise linear or non-linear. 
    \label{fig:RETschematic}}
\end{SCfigure}

Then optimising the choice of $\btau$  at $\s$ optimises over the bounded constrained set $E_f(\s(\y)) \ge \s$. This is a simple consequence of convexity.   
As detailed in Section~\ref{sec:ExplFamily} below, 
 the minimised KL metric $\kappa(\s)$ is a convex function of $\s$,  and hence
 $\kappa(\s_1)>\kappa(\s)>\kappa(\s_{\bzero})=0$ for any $\s_1>\s>\s_{\bzero}$ elementwise.   Hence  the ET optimal vector $\btau$ at target $\s$ minimises the KL divergence over all possible target scores $\s_1>\s.$   Figure~\ref{fig:RETschematic} highlights this with a schematic emphasising the exploration of modest perturbations of the baseline using the ET approach.

\subsection{ET and MaxEnt} 

The KL divergence that ET minimises is 
 $$ {KL}_{p|f} = \int_\y \log(f(\y)/p(\y))f(\y)d\y = H_{p|f} - H_f $$
 where 
  $H_f = -  \int_\y \log(f(\y))f(\y)d\y$ is the entropy (also known as the differential entropy) of $f(\y)$ and 
  $H_{p|f} = -  \int_\y \log(p(\y))f(\y)d\y$ is the cross-entropy of $p(\y)$ under $f(\y).$    Thus ${KL}_{p|f}$  measures entropy in $p(\y)$ {\em relative to} the absolute entropy $H_f$ of $f(\y)$.   In the 1960s onwards for several decades, so-called \lq\lq objective Bayesian'' developments included a main theme in scientific research pioneered by statistical physicists that defined the {\em maximum entropy (MaxEnt)} approach to specifying priors based on sets of constraints taken as partial prior information.  The seminal work is by Edwin~T.~Jaynes~\citep[][available at~\href{http://bayes.wustl.edu/etj/articles/theory.1.pdf}{http://bayes.wustl.edu/etj/articles/theory.1.pdf}]{Jaynes1957} and MaxEnt has a long history since then at the interface between statistics, information theory and statistical physics. 
The basic idea underlying MaxEnt  is to choose $f(\y)$ to satisfy some set of constraints, but otherwise to be \lq\lq least informative'' in the sense of maximizing information-theoretic uncertainty characterized explicitly via entropy. 
 
MaxEnt  can be viewed as special cases of ET in which the baseline distribution has a \lq\lq flat''  p.d.f. $p(\y),$ i.e., a limiting (possibly improper) \lq\lq uninformative'' baseline, so that $H_{p|f}$ is constant for any $f(\y).$  In such cases, minimizing ${KL}_{p|f}$ is equivalent to maximizing $ H_f$, again subject to the constraints.   Hence this specialization of ET reproduces MaxEnt.      The utility of MaxEnt as a constructive approach to defining distributions based on sets of expectations is clearly limited, as this result shows.  The requirement of a uniform baseline $p(\y)$ leads to the same limitations and issues as encountered in the historical objective Bayesian (fruitless) search for globally uniformative priors (notable among which is the fact that uniformity is transformation dependent, in general).

\section{ET Theory: Proofs and More Insights} 
\subsection{Direct Proof of ET Solution: Discrete Cases} 

The proof is easy and direct using Lagrange multipliers. Here $\y$ has support $\y \in \{ \y_1,\ldots,\y_n\}$ with baseline  p.d.f. $p_i = p(\y_i)$ and any  p.d.f. that satisfies the constraints $f_i = f(\y_i)$ over $i=\seq 1n.$  Write $\p = (p_1,\ldots,p_n)'$ and $\f = (f_1,\ldots,f_n)'.$ The ET problem is to choose $\f$ and $\btau$ to minimise
$$ R(\f,\btau) = \sum_{i=1{:}n} \log(f_i/p_i) f_i  + \sum_{j=1{:}q}\tau_j \{ s_j- \sum_{i=1{:}n} s_j(\y_i) f_i \}$$
where $\btau = (\tau_1,\ldots,\tau_q)'$ is a vector of Lagrange multipliers that enforce the constraints.

Differentiate with respect to any $f_i$ and equate to zero to give 
$$0 = \frac{\partial R(\f,\btau)}{\partial f_i}  = 1+ \log(f_i/p_i)  - \sum_{j=1{:}q}\tau_j s_j(\y_i)  = 
1+ \log(f_i/p_i)  -\btau'\s(\y_i) $$
so that  
$\log(f_i)  =  \log(p_i)  -1 +  \btau'\s(\y_i). $
Differentiating $R(\f,\btau)$ a second time yields  the Hessian matrix that is diagonal with $i^{th}$ diagonal element $1/f_i$; this is positive definite ensuring a minimum.  
This then results in the minimizing solution 
$$f_i  \propto  p_i \exp(\btau'\s(\y_i)),\ i=\seq 1n, $$   also showing that the \lq\lq tilting'' parameter vector $\btau$ is the vector of Lagrange multipliers. 

 The above proof can be replayed with $\s(\y),\s,\btau$ including elements $s_1(\y)=s_1=1$ and $\tau_1$ that explicitly serves to define normalization of $f(\cdot).$ This is not necessary, however, as the above proof defines the ET solution up to proportionality, and normalization follows.

\subsection{Proof in Continuous and General Cases: The Informal Proof}

The obvious limiting argument defines the solution in continuous cases based on the above, easily-derived solution in the discrete case.  Take $n\to\infty$ with $p_i \to p(\y_i)\delta \y_i$ and $f_i \to f(\y_i)\delta \y_i$ for differentials $\delta \y_i.$  Then any sum with weights $f_i$ becomes an integral with respect to $f(\y)\delta\y,$ so the ET constraints are general expectations.   The general solution then trivially follows as the limiting form of the discrete case above, $f(\y)\delta\y \propto p(\y)\exp(\btau'\s(\y))\delta\y$ for all $\y.$ 
 
\subsection{General Mathematical/First-Principles Proof}
 
Formal functional calculus~\citep{FoxFunctionalCalculus1987} delivers a rigourous mathematical proof.   Whatever the sample space, let $a(f)$ be any functional of the  p.d.f. $f(\y).$ 
Consider the functional derivative of $a(f)$ with respect to the specific value $f(\y)$ at any chosen $\y$ in the support of $f(\cdot)$, namely
$$\frac{\partial a(f)}{\partial f(\y)} = \lim_{\epsilon\to 0} \frac{A_{f,\epsilon}(\y)}{\epsilon} \quad\textrm{where}\quad A_{f,\epsilon}(\y) =  a(f+\epsilon\delta_{\y}) - a(f) $$ 
and where $\delta_{\y}$ is the point-mass at $\y,$ i.e., the Dirac delta function  $\delta_{\y}(\y) = 1$ if $\y=\y$ and 0 otherwise.  
For small $\epsilon$ the function $f(\y)+\epsilon\delta_{\y}(\y)$ is a local perturbation of $f(\y),$ corresponding to the concept and definition of standard derivatives of functions.

Examples of interest here are as follows. 
\bn
\i {\blu Expected Score.}  Suppose $a(f) = \int_\y s(\y)f(\y)d\y$ for a specified scalar score function $s(\y)$ that does not depend on $f(\cdot).$   Then
$$
A_{f,\epsilon}(\y)   = \int_\y s(\y) \left\{ f(\y)+\epsilon \delta_{\y}(\y) \right\} d\y  - \int_\y s(\y) f(\y) d\y 
 = \epsilon \int_\y s(\y) \delta_{\y}(\y) d\y  = \epsilon s(\y)\\
$$
so that 
$$\frac{\partial a(f)}{\partial f(\y)} = s(\y).$$ 
\i {\blu Entropy.}  Suppose $a(f) = \int_\y \log(f(\y))f(\y)d\y$.    Then
$$A_{f,\epsilon}(\y)   = \int_\y \left\{ \log \left\{ f(\y)+\epsilon \delta_{\y}(\y) \right\} \left\{ f(\y)+\epsilon \delta_{\y}(\y) \right\}   - \log(f(\y)) f(\y) \right\}d\y.$$
Now, for any $f$ and small $z$ note that $\log(f+z) = \log(f) + z/(z+f) + o(z^2)$ so that the integrand defining $A_{f,\epsilon}(\y)$ is 
\beas
&&  \left\{ \log(f(\y))+ \frac{\epsilon \delta_{\y}(\y)}{f(\y)+\epsilon \delta_{\y}(\y)} + o(\epsilon^2) \right\} \left\{ f(\y)+\epsilon \delta_{\y}(\y)\right\} - \log(\y) f(\y) \\
&=& \epsilon \left\{ \log(f(\y)) +f(\y)/(f(\y)+\epsilon\delta_{\y}(\y))\right\}  \delta_{\y}(\y)  + o(\epsilon^2). 
\eeas
On integration this gives $A_{f,\epsilon}(\y) = \epsilon \left\{ \log(f(\y)) +f(\y)/(f(\y)+\epsilon) \right\}  + o(\epsilon^2)$  and so
$$\frac{\partial a(f)}{\partial f(\y)} =\log(f(\y))+1.$$  
\en
The above examples underlie the formal proof of the ET result.   The goal is to find $f(\y)$ to minimise 
$$ R(f,\btau) = \int_\y \log(f(\y)/p(\y)) f(\y)d\y  + \sum_{j=1{:}q}\tau_j \{ s_j- \int_\y s_j(\y) f(\y)d\y \}$$
where $\btau = (\tau_1,\ldots,\tau_q)'$ is a vector of Lagrange multipliers that enforce the constraints. 
Functional differentiation with respect to $f(\y)$ at any value $\y$ now exploits the results of the above examples and defines the optimizing p.d.f.  Explicitly, 
$$  \frac{\partial R(f,\btau)}{\partial f(\y)} =  \log(f(\y)) + 1 - \log(p(\y)) - \btau'\s(\y)  =  1+ \log(f(\y)/p(\y))  - \btau'\s(\y).$$
Note that this is the direct general extension of the form above in the discrete case.   Then, 
equating this functional derivative to zero yields the ET result 
$f(\y) \propto p(\y) \exp(\btau'\s(\y)).$  Further, it is easily confirmed that the functional second derivative of $R(f,\btau)$ with respect to $f(\y)$ at any $\y$ is just 
$1/f(\y)$; this is positive everywhere on the support hence the solution is a minimum.   

\subsection{General Proof via Euler–Lagrange}

The general proof above is from first-principles and is instructive in terms of understanding functional calculus.  The result can alternatively be directly deduced from (a special case of) the Euler–Lagrange (E-L) equation for functional derivatives that builds on such first-principles~\citep{FoxFunctionalCalculus1987}. The relevant E-L theory is as follows. 
Suppose $H(\y,z)$ is a real-valued function of the vector $\y$ and a scalar $z$ and that $H(\cdot,z)$ is differentiable in $z$. Consider a functional of the p.d.f. $f(\cdot)$ of the form 
$a(f) = \int_{\y} H(\y,f(\y))d\y$  where the integral is assumed to exist.  Then the functional derivative of  $a(f)$ with respect to a point value $f(\y)$ at any given $\y$ is  
$$\frac{\partial a(f)}{\partial f(\y)} = h(\y,f(\y))\qquad \textrm{where}\qquad h(\y,z)=\frac{\partial H(\y,z)}{\partial  z}.$$
Now take $z>0$ and note the following. 
\bi
\i The choice $H(\y,z) = z \log(z/p(\y)) + \btau'\s - z\btau'\s(\y)$ implies that $a(f) = R(f,\btau),$   and $h(\y,z)= 1+ \log(z) -\btau'\s(\y)$.
\i The choice $H(\y,z) = 1+ \log(z) -\btau'\s(\y)$ has derivative $h(\y,z)=1/z.$ 
\ei
These observations deliver the above ET result directly.  Note that, as in the discrete case, ensuring normalisation might use elements $s_1(\y)=s_1=1$ and $\tau_1$, though this is not necessary.

\subsection{The Most General, Statistical, Intuitive and Contextual Proof}
 
The really insightful proof that covers all cases is as follows. Consider {\em any} distribution with p.d.f. $g(\y)$ that satisfies the ET constraint $E_g(\s(\y))=\s,$ i.e., focus only on these relevant distributions. One of these is $f(\y)$ defined above for the vector $\btau$ implied by the chosen target $\s,$ but there are many other distributions consistent with that target expectation vector.   Now note that, for any such p.d.f. $g(\y),$ it is immediate that 
\begin{eqnarray*}
{KL}_{p|g} &=& \int_\y \log(g(\y)/p(\y))g(\y)d\y  \\
&=& \int_\y \log(g(\y)/f(\y))  g(\y)d\y  + \int_\y \log(f(\y)/p(\y))  g(\y)d\y \\
&=& {KL}_{f|g} + \int_\y \{ \btau'\s(\y)-c(\btau) \}g(\y)d\y \\
&=& {KL}_{f|g} + \btau'\s-c(\btau)  \\
&=& {KL}_{f|g} + {KL}_{p|f}
\end{eqnarray*}
with the last step following since ${KL}_{p|f} = \kappa(\s) = \btau'\s-c(\btau)$  from eqn.~(\ref{eq:KLpfs}).
Now, ${KL}_{f|g}$ is non-negative and positive unless  $g(\y) = f(\y).$ 
Hence ${KL}_{p|g} \ge {KL}_{p|f}$ with equality only when $g(\y) = f(\y).$   This proves the result: among {\em all} distributions satisfying the expectation constraint defined by the target $\s,$ the ET p.d.f. $f(\y)$ minimises the KL divergence.

 \section{Exponential Family Theory \label{sec:ExplFamily}}

\subsection{ET Defines Exponential Families} 

For any chosen target $\s$, if an ET solution in $\btau$ exists, it is unique. Reciprocally, any feasible choice of $\btau$ implies the corresponding $\s,$ so that any choice of $\btau$ defines the p.d.f. in  \eqn{fETpdf} with the implied $\s = \s(\btau).$    As $\s$, and hence $\btau,$ varies it is of theoretical interest to note that the analysis defines-- as has already been noted-- a class of exponential family distributions with p.d.f.s in \eqn{fETEXPF}.  
Some comments and theoretical points arising for this defined family of distributions are noted~\citep[and see][for general background on exponential families, convexity and other theoretical aspects]{BarndorffNielsen1978}.

Assume that $\y \sim f(\y)$ as defined in \eqn{fETEXPF} for a given vector of parameters $\btau.$ As $\btau$ varies, this defines a family of distributions representing entropically tilted variations of the baseline $p(\y).$  In exponential family terminology, $\btau$ is the natural parameter of the distribution and 
$c(\btau)$ is  known as the {\em cumulant function} of the parameter vector $\btau.$  The analysis here relates to {\em regular} exponential families in which (i)  the support of $f(\y)$ does not depend on parameters $\btau;$
(ii) the elements of $\btau$ are continuous and $c(\btau)$ is at least twice-differentiable in $\btau$; 
and (iii) the $\btau$ parameters are not subject to constraints (although this  can be relaxed). 

\newpage

\subsection{Convexity in Tilting Parameters} 
In regular exponential families, the parameter space for $\btau$  is a convex set and $c(\btau)$ is a convex function from well-known theory of regular exponential families~\citep{BarndorffNielsen1978}.  

The standard proof is as follows.  Take any two values $\btau_1, \btau_2$ with corresponding p.d.f.s 
$f_j(\y) =  p(\y) \exp(\btau_j'\s(\y) - c(\btau_j))$ and $ \exp(c(\btau_j)) = \int_\y  \exp(\btau_j'\s(\y)) p(\y)d\y $  for $j=1,2.$  
Consider  a  third distribution with parameter $a_1\btau_1+a_2\btau_2$ for any $a_1\in (0,1)$ and $a_2=1-a_1.$  Then
$$\exp\{c(a_1\btau_1+a_2\btau_2)\} = \int_\y  \exp\{(a_1\btau_1+a_2 \btau_2)'\s(\y)\} p(\y)d\y = \int_\y g_1(\y)g_2(\y) d\y$$
where  $g_j(\y) = \{\exp(\btau_j'\s(\y)) p(\y)\}^{a_j}$ for $j=1,2$.
Now recall H{\"o}lder's  inequality: for 
any two functions $g_j(\y)>0$, $j=1,2,$ for which the integrals exist and any $a_1,a_2$ as above, 
$$\int_\y g_1(\y)g_2(\y)d\y \le \left\{\int_x g_1(\y)^{1/a_1}d\y \right\}^{a_1} \left\{\int_x g_1(\y)^{1/a_2}d\y \right\}^{a_2}.$$
This leads immediately to  
$\exp\{c(a_1\btau_1+a_2\btau_2)\} \le \exp\{a_1c(\btau_1)\} \exp\{a_2c(\btau_2)\}$  which, on taking logs, yields
$ c(a_1\btau_1+a_2\btau_2) \le a_1c(\btau_1) + a_2c(\btau_2)$  so that $c(\cdot)$ is convex.  

\subsection{Some Moments of Score Functions} 

The derivative (vector) of $c(\btau)$ defines the mean vector of $\s(\y),$ i.e., 
\begin{equation}\label{eq:Esytau}\s  \equiv E_f( \s(\y) ) = \frac{\partial c(\btau)}{\partial\btau}.\end{equation}  
Since $c(\btau)$ is convex,  each element $s_i$ of $\s$ is an increasing function of the corresponding element $\tau_i$ of $\btau$, and vice-versa.
Note, however and importantly, an element $s_i$  may or may not be increasing in $\tau_j$ for $j\ne i.$
 
The second derivative (matrix) of $c(\btau)$ defines the variance matrix of $\s(\y),$ i.e., 
\begin{equation}\label{eq:Vsytau} V_f(\s(\y)) =\frac{\partial^2c(\btau)}{\partial\btau\partial\btau'}.\end{equation}
That this is positive definite for all $\btau$ is implied by, and implies, the convexity of  $\c(\btau).$

\subsection{Reciprocal Relationships for Tilting Parameters} 

It also follows that $$ \frac{\partial\btau}{\partial\s'} = V_f(\s(\y))^{-1}.$$
To prove this, simply observe that $$\I = \frac{\partial\s}{\partial\s'} = \left\{ \frac{\partial\s}{\partial\btau'}\right\} \frac{\partial\btau}{\partial\s'} 
 = \left\{ \frac{\partial^2 c(\btau)}{\partial\btau\partial\btau'}\right\} \frac{\partial\btau}{\partial\s'}
  = V_f(\s(\y))  \frac{\partial\btau}{\partial\s'}
$$ using the relationships of  eqns.~(\ref{eq:Esytau}) and (\ref{eq:Vsytau}); the result follows.

Recall from above that the KL divergence of $p(\cdot)$ from $f(\cdot)$ is 
${KL}_{p|f} = \kappa(\s) = \btau'\s - c(\btau).$    Now consider how this varies with chosen target $\s$ via its implied value of $\btau.$  Note that 
\begin{equation}\label{eq:taueqderivks}
 \frac{\partial \kappa(\s)}{\partial\s} = \btau +  \frac{\partial\btau'}{\partial\s\phantom{'}}\s -   \frac{\partial\btau}{\partial\s}'\frac{\partial c(\btau)}{\partial\btau}
= \btau + \frac{\partial\btau'}{\partial\s\phantom{'}} \left\{ \s -  \frac{\partial c(\btau)}{\partial\btau} \right\} = \btau.
\end{equation}
That is, the parameter vector $\btau$ defined by any specified target $\s$ is in fact just the gradient vector of the KL divergence at that target.

\subsection{Convexity of KL Metric} 

Using the previous two results it further follows that
\begin{equation}\label{eq:taueq2ndderivks} \frac{\partial^2\kappa(\s)}{\partial\s\partial\s'} = \frac{\partial\btau}{\partial\s'} = V_f(\s(\y))^{-1}
\end{equation}
which is positive definite for all $\s,$ implying the important property that $\kappa(\s)$ is a convex function of $\s.$

The identities of eqns.~(\ref{eq:taueqderivks})~and~(\ref{eq:taueq2ndderivks})  are apparently not so well-known or elaborated in the literature. They highlight
theoretical links between the cumulant function $c(\btau)$  and the KL metric $\kappa(\s)$ of the exponential family that are related to the ability to 
parametrise the exponential family in either $\btau$ or $\s.$ Specifically: 
\bi
\i Note that~\eqn{taueqderivks} is a dual of~\eqn{Esytau};  $\s$ is defined by $\btau$ as the derivative of the cumulant function $c(\btau)$, while reciprocally $\btau$ is defined by $\s$ as the derivative of the KL metric $\kappa(\s)$. 
\i  Note further that~\eqn{taueq2ndderivks} is a dual of~\eqn{Vsytau};  the variance matrix of  $\s(\y)$ is defined by $\btau$ as the second derivative of the cumulant function $c(\btau)$, while reciprocally the precision matrix (inverse of the variance matrix) is defined by the second derivative of the KL metric $\kappa(\s)$.
\ei
  
\subsection{Small Perturbations of Baseline} 
In connection with the perspective of using ET to explore \lq\lq small perturbations'' of the baseline $p(\y)$ using $\s = \s_{\bzero}+ \bepsilon$ for some \lq\lq small'' increment $\bepsilon$ to the baseline expected score, the above results deliver first-order approximations as follows.   Write $\V = V_p(\s(\y)) \equiv V_f(\s(\y))$ when $f(\cdot)\equiv p(\cdot)$ corresponding to $\btau = \bzero,$  and let $\bLambda = \V^{-1}$ be the corresponding precision matrix of the score vector under the baseline distribution. 

Consider a first-order Taylor series approximation to   $\btau = \btau(\s)$ as a function of $\s$, relevant for very small $\bepsilon.$  Exploiting the results above yields the approximation $\btau \approx \bLambda\bepsilon.$  The dual approximation to $\s = \s(\btau)$ as a function of $\btau$ for  very small $\bepsilon$ is $\s \approx \s_{\bzero} + \V\btau.$  These results show how the baseline dependencies among the elements of the chosen score vector play roles in influencing tilting for a given target score, and reciprocally in defining implied targets based on choices of tilting parameters.  

It is easily verified that such small perturbations yield the corresponding local approximation to the KL metric given by $\kappa(\s) \approx (\s-\s_0)'\V^{-1}(\s-\s_0) = \btau'\V\btau/2.$

 \newpage
 
\section{Theoretical Examples with Moment Constraints \label{sec:TExamples}} 
 
Two very simple, idea-fixing theoretical examples are as follows. 

\subsection{Poisson Mean Example \label{sec:egPoissonmean}} 
Suppose $\y=y$ is scalar and $p(y)$ is Poisson with mean $\mu.$ Take $\s(\y)= s(y)=y$ with target $\s=s.$ Then $f(y)$ is Poisson with mean 
$s=\mu\exp(\tau)$ where $\btau=\tau = \log(s/\mu)$.  Here 
$c(\tau) = \mu(\exp(\tau)-1) = s-\mu$, which is convex in $\tau$, of course, and also in $s$.\footnote{$c(\tau)=s-\mu$ is linear in $s$ with zero second derivative, so is on the boundary of convexity. Under the usual definition of convexity as used above in which bounds are based on $\le$, linear functions are both convex and concave. However, using strict convexity in which the bounds are strictly $<,$ a linear function is not convex.} 
Further, $\kappa(s) =\mu+ (\tau-1)\mu\exp(\tau) = \mu+ s\log(s/\mu)-s$, which is convex in $s$, of course, but {\em not} convex in $\tau$.

\subsection{Bivariate Normal Mean Example \label{sec:egbivarnormalmean}} 
Suppose $p(\y)$ is bivariate normal $\y\sim N(\bzero,\V)$ with 
$$\V = \begin{pmatrix} 1&\rho\\ \rho&1 \end{pmatrix}\qquad\textrm{and}\qquad \bLambda=\V^{-1} = (1-\rho^2)^{-1}\begin{pmatrix} 1&-\rho\\ -\rho&1\end{pmatrix}$$ for some correlation $\rho.$   Take a $q=2-$dimensional score vector $\s(\y)=\y$ so that $\s_{\bzero}=E_p(\s(\y))=\bzero$. It trivially follows that  $f(\y)$ is bivariate normal $N(\s,\V)$ with
 target score $\s =\V\btau$ and thus $\btau = \bLambda\s.$ Also  $c(\btau)=\btau'\V\btau/2$ and the minimised KL metric is $\kappa(\s) = \btau'\V\btau/2= \s'\bLambda\s/2.$   
 
Suppose the target score $\s\ge \bzero$ with at least one positive element; i.e., the target reflects an increase in one or both dimensions relative to the baseline $\s_{\bzero}=\bzero.$  The optimising $\btau$ vector is
$$ \begin{pmatrix}\tau_1\\ \tau_2\end{pmatrix}  = (1-\rho^2)^{-1} \begin{pmatrix} s_1-\rho s_2\\ s_2-\rho s_1\end{pmatrix}. $$ 
In cases when $\rho\le 0,$ $\tau_1$ and $\tau_2$ are non-negative and at least one of them is positive since one or both of $s_1$ and $s_2$ are positive. In cases when $\rho>0,$ one of the $\tau_j$ may be negative depending on the values of target scores $\s\ge \bzero.$  Specifically, $\tau_1\ge 0,\tau_2\le 0$ if $\rho s_1\ge s_2\ge 0,$ while  $\tau_1\le 0,\tau_2\ge 0$ if $\rho s_2\ge s_1\ge 0,$ and $\tau_1\ge 0,\tau_2\ge 0$ if $\rho s_1\le s_2\le s_1/\rho.$  

Reciprocally,  values of $\btau$ that satisfy $\tau_2 \ge \textrm{max} \{ -\rho\tau_1, -\tau_1/\rho \}$ imply expected scores $\s\ge \s_{\bzero}= \bzero;$ 
this defines the RET schematic shown earlier in Figure~\ref{fig:RETschematic}.

On a technical point, recall the general results that $c(\btau)$ is a convex function of $\btau$ while $\kappa(\s)$ is a convex function of $\s.$ This specific example shows that $c(\btau)$ can also be convex in $\s$ at the optimized $\btau=\btau(\s),$ and $\kappa(\s)$ can also be convex in $\btau$ under that relationship.  This is interesting to point out,   but is not true in general as the univariate Poisson mean example of Section~\ref{sec:egPoissonmean} above shows. 
 
 \newpage
 
 \section{Practicable Examples of ET with Marginal Quantile Constraints \label{sec:QExamples}} 
  
A  far-reaching context involves conditioning on a set of specified target quantiles for the univariate marginal distribution of a chosen function or set of functions of $\y$.   Constraining quantiles is  natural from an application viewpoint, as quantiles are often a main focus in elicitation of the role of external information.  Quantile scores also always lead to a defined solution to the ET optimization problem, while other choices of score may not. For example, if some elements of $\s(\y)$ are linear in $\y$ then a solution exists only if the moment generating function of $p(\y)$ exists; this rules out many practical settings such as common models delivering predictive distributions that are  multivariate log T~\citep[e.g. appendix in][]{West2021decisionconstraints}. 

\subsection{Quantile Specifications} 

For example here, take $\phi(\y)$ as a univariate function such as: 
\bi
\i $\phi(\y) = \bone'\y$ to constrain and update predictions of $\y$ based on a specified set of target quantiles for the resulting distribution of the sum of elements of $\y.$ 
\i $\phi(\y) = y_1$ to update  based on target quantiles for the marginal 
of the first element of $\y.$ 
\i  $\phi(\y) = y_2/y_1$ to update  based on target quantiles for the marginal 
of the relative value of the second to first element of $\y.$ 
\ei 
The generality is evident, as are extensions to quantiles of multiple functions.  Continuing in this first  example, quantile constraints are implemented as follows. 
\bi
\i Choose a vector of target quantiles for $\phi(\y)$ in the $q-$vector $\f = (f_1,\ldots,f_q)'$ with $f_{i-1}\le f_i$ for $i=2{:}q.$ 
Extend the notation to define and include 
endpoints $f_0=-\infty$ and $f_{q+1}=\infty$, giving intervals 
 ${\cal I}_i = (f_{i-1},f_i]$  for $i=\seq 1{(q+1)}.$ 
\i  Take $\s(\y) = ( s_1(\y),\ldots,s_q(\y))'$ with elements $i=\seq 1q$ given by 
 $s_i(\y) = I(\phi(\y) \in {\cal I}_i ) $ where  $I(\cdot)$ is the indicator function. 
\i  Fix the target score vector $\s = (s_1,\ldots,s_q)'$ to define probabilities between the quantiles. Require that $s_i>0$ and that $\bone'\s<1,$ naturally, 
and set $s_{q+1} = 1-\bone'\s >0.$ 
 For example,  $s_i = q/(q+1)$ for each $i=\seq 1q$
defines equal quantile regions:  $q=3$ implies the constrained distribution of $\phi(\y)$ has lower quartile, median and upper quartile  in $\f = (f_1,f_2,f_3)',$  
while taking $q=9$ constrains the deciles.    In these cases, 
the target  constraints impose $\f$ as the endpoints of the $100/(q+1)\%$ probability intervals in the marginal distribution of $\phi(\y).$  
\i Taking unequal probabilities provides generality; for example, $\s = (0.25,0.25,0.45)'$ constrains the lower quartile, median and upper 95\% point of the distribution, and so forth. 
\i Taking a very large value of $q$ means that the specified $\f$ vector  maps out an increasingly accurate representation of the full target marginal distribution of $\phi(\y)$. That is, this is a Bayesian decision-analytic approach to updating $p(\y)$ to $f(\y)$ that conditions on a modified marginal distribution for $\phi(\y).$ Extensions to multivariate margins are immediate. 
\ei

\subsection{Almost Exact/Deterministic Constraints using Quantile ET}  

A really interesting and important use of quantile ET is to explore and potentially condition upon \lq\lq almost deterministic'' constraints. This was first identified, explored and exploited in~\cite{West2021decisionconstraints} as one approach to conditioning forecast distributions on almost-exact constraints. 
The idea is simple:  take two \lq\lq extreme'' quantiles for the predictive distribution of a chosen score function, and use ET to explore and/or enforce the target that the probability under $f(\y)$ between the two extremes in that score dimension is \lq\lq very small''. 

\subsection{Analytic Solution under Quantile Constraints}  

Perhaps somewhat surprisingly, the ET optimization can be evaluated analytically in this setting of quantile constraints-- in spite of the generality of examples (different sets of functions $\phi(\y)$).  This new theoretical result derives from the fact that the score vector $\s(\y)$ is a vector of indicator functions, so knowing the mean of this vector under $p(\y)$ (the set of prior probabilities on each of the constrained regions) is enough to define the analytic solution for  $\btau$.
 
For each $i=\seq 1q,$ define $s_{0,i} = E_p[ s_i(\y) ]$, the initial probabilities under $p(\y)$ on each of the $k$ intervals ${\cal I}_i$.     Write $\s_{\bzero} = (s_{0,1},\ldots,s_{0,q})'$ and define  $s_{0,q+1} = 1-\bone'\s_{\bzero} = 1-\sum_{i=\seq 1q} s_{0,i}$ as the probability on ${\cal I}_{q+1}$ under $p(\y).$    Require $s_{0,i}>0$ for $i=\seq 1{(q+1)},$ naturally.

\paragraph{\em\blu Quantile Constraint Result:} 
The $q-$vector $\btau = (\tau_1,\ldots,\tau_q)'$ has elements 
\begin{equation} \label{eq:kuantileconstraintresult} 
\tau_i = \log\{ (s_{0,q+1} s_i )/(s_{q+1} s_{0,i}) \}, \qquad i=\seq 1q.
\end{equation} 
The proof is as follows. 
From the ET defining \eqn{ETtau}, note that, for each $i=\seq 1q,$ 
\beas
 0 &=& \int_{\y} \{s_i(\y) - s_i \} \e^{\btau'\s(\y)} p(\y)d\y  
  =  \sum_{j=\seq 1{(q+1)}} \int_{\y: \phi(\y)\in {\cal I}_j} \{s_i(\y) - s_i \} \e^{\btau'\s(\y)} p(\y)d\y \\
    &=& (1-s_i)\e^{\tau_i} s_{0,i}  -s_i \sum_{j=\seq 1q, j\ne i}  \e^{\tau_j} s_{0,j} - s_i s_{0,q+1} \\
    &=&  \e^{\tau_i} s_{0,i}  -s_i \sum_{j=\seq 1q}  \e^{\tau_j} s_{0,j} - s_i s_{0,q+1}
 \eeas
 using the facts that $\btau'\s(\y) = \sum_{j=\seq 1q} \tau_j  I( \phi(\y)\in {\cal I}_j) $ which 
 reduces to $\tau_j$ when $\phi(\y)\in {\cal I}_j$ for any $j=\seq 1q,$ and is zero when  $\phi(\y)\in {\cal I}_{q+1}$.

 For each $i=\seq 1q$, let $\theta_i = \exp(\tau_i)$ and set $\btheta = (\theta_1,\ldots,\theta_q)'.$   Further, 
 let $\S_0$ be the $q\times q$ diagonal matrix   $\S_0 = \diag(\s_{\bzero}).$ 
 The above set of equations gives the vector form 
  $ (\S_0  - \s \s_{\bzero}') \btheta = s_{0,q+1} \s  $ with solution $$\btheta = s_{0,q+1} (\S_0  - \s \s_{\bzero}')^{-1} \s.$$
  Now
  $$  (\S_0  - \s \s_{\bzero}')^{-1} = \{ (\I - \s \s_{\bzero}'\S_0^{-1}) \S_0\}^{-1} = \S_0^{-1} (\I - \s \bone')^{-1} 
   = \S_0^{-1} (\I + \s \bone'/s_{q+1})$$
  using the fact that $\S_0^{-1}\s_{\bzero} = \bone$ and standard linear algebra results for matrix inversion.    
This result is valid as  (i) the diagonal matrix $\S_0$ is invertible since its diagonal elements $s_{0,j}$ are all positive, and (ii)
the matrix  $\I - \s \bone'$ is invertible since it has determinant $1-\bone'\s = s_{q+1}>0.$   As a result, 
  $$\btheta = s_{0,q+1} \S_0^{-1} (\I + \s \bone'/s_{q+1})  \s = s_{0,q+1} \S_0^{-1} \s (1 + (1-s_{q+1})/s_{q+1}) 
  =  (s_{0,q+1}/s_{q+1})  \S_0^{-1} \s.  $$
   Elementwise, $\theta_i =   (s_{0,q+1}/s_{q+1}) (s_i/s_{0,i})$ and the result follows since $\tau_i = \log(\theta_i)$ for each $i=\seq 1q.$

  It is also easily shown that the inverse transformation  defining $\s$ in terms of $\btau$ as in eqn.~(\ref{eq:Esytau}) is given elementwise by 
 $$s_i = s_{0,i}\e^{\tau_i -c(\btau)}, \qquad  i=\seq 1q,$$ where 
$$ c(\btau)=
 \log\{ s_{0,q+1} + \sum_{j=\seq 1q} s_{0,j}\e^{\tau_j} \}
 = \log\{1+ \sum_{j=\seq 1q} s_{0,j}(e^{\tau_j}-1)\}$$
 is the cumulant function in this setting.   
Note that $s_i$ is strictly increasing in $\tau_i$, but strictly {\em decreasing} in $\tau_j$ for $j\ne i.$  There are specific choices of $\s$  that define optimal $\btau$ vectors with strictly positive elements, but others for which some elements of $\btau$ are negative.

\subsection{Synthetic Examples: Quantile ET Relevant to Bayesian Forecasting}

\begin{figure}[htp!]
\centering
\hskip-.15in\includegraphics[width=3.4in]{\figdir/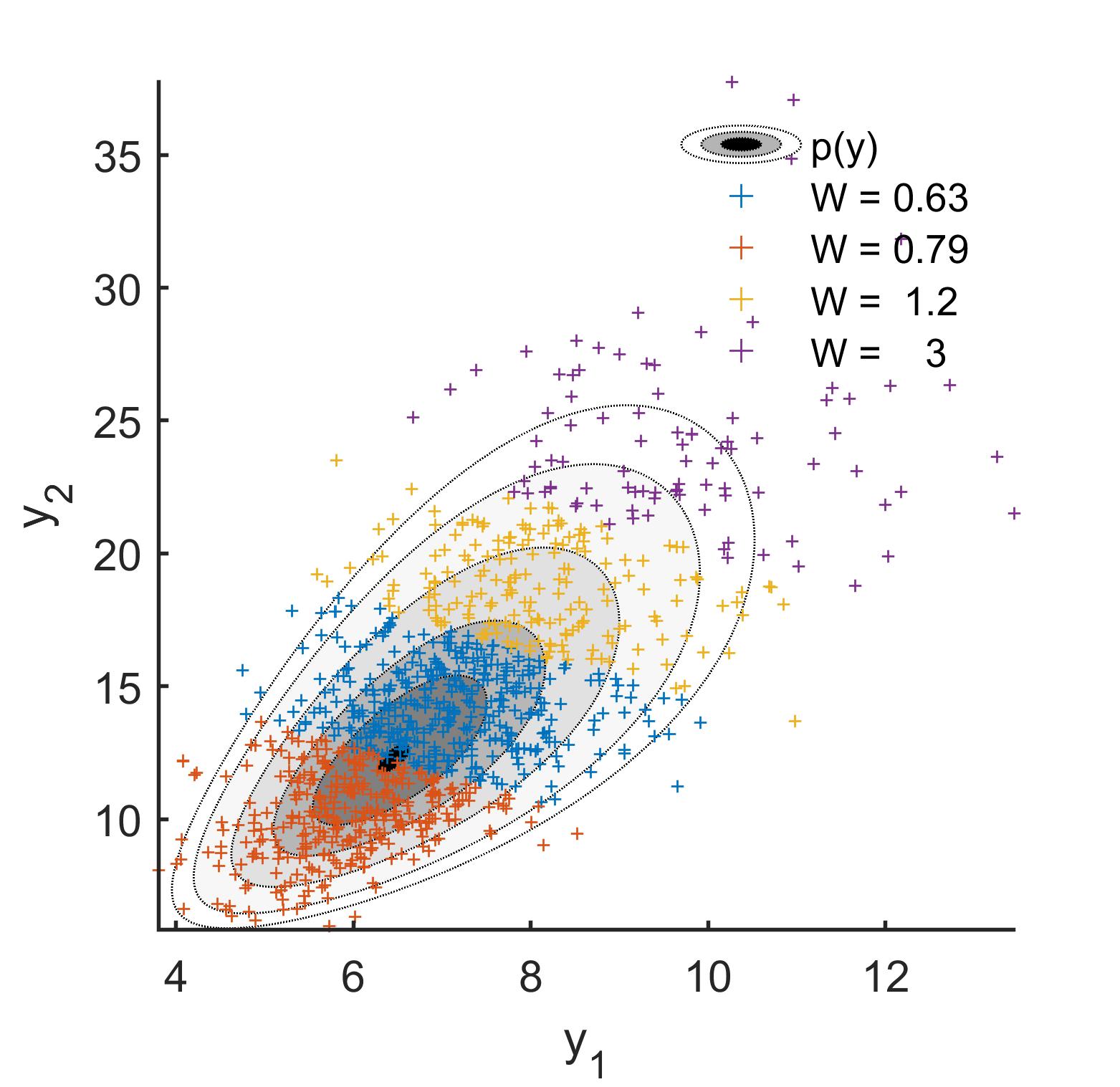} 
\hskip-.25in\includegraphics[width=3.4in]{\figdir/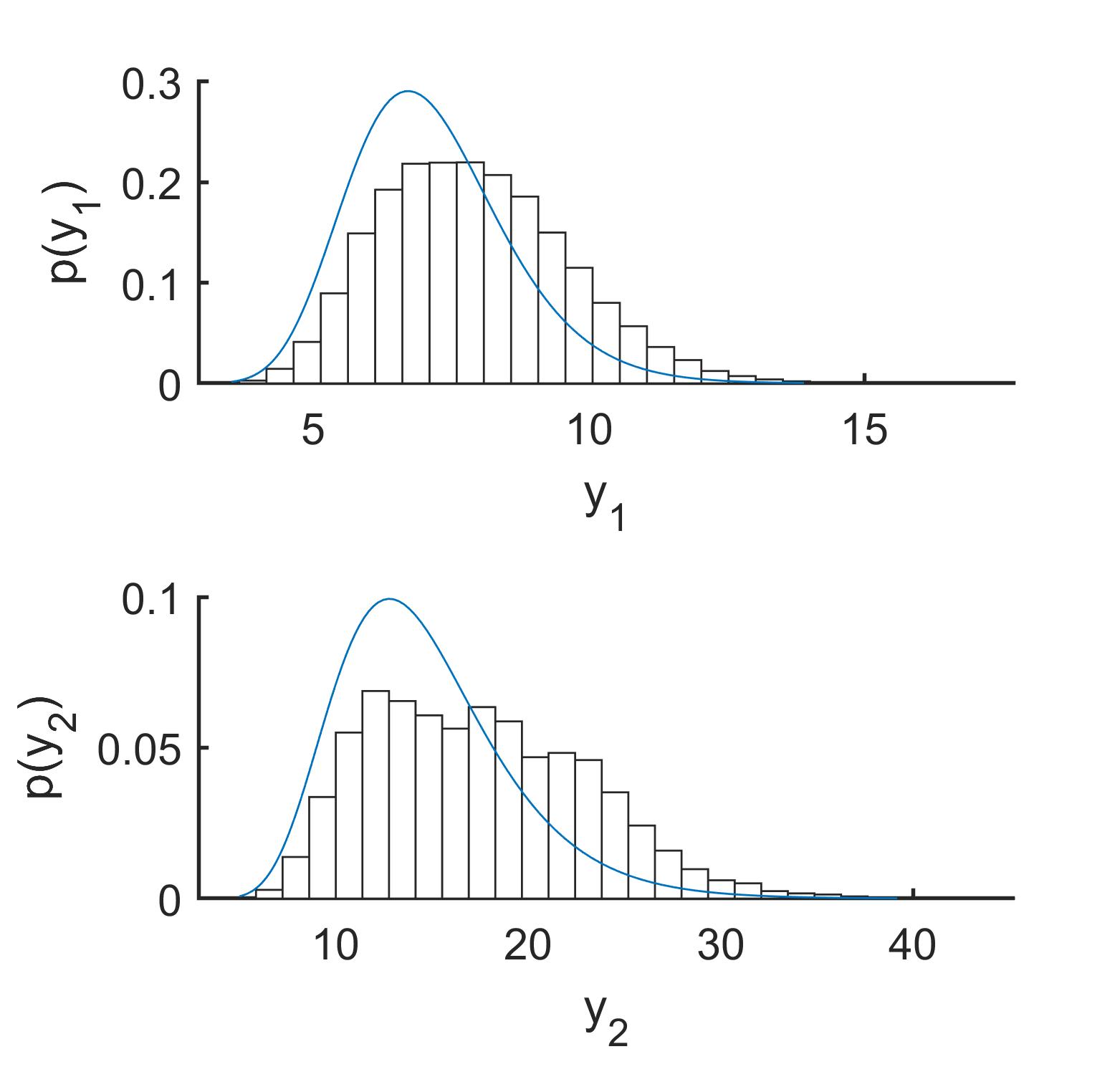} \\
\hskip-.15in\includegraphics[width=3.4in]{\figdir/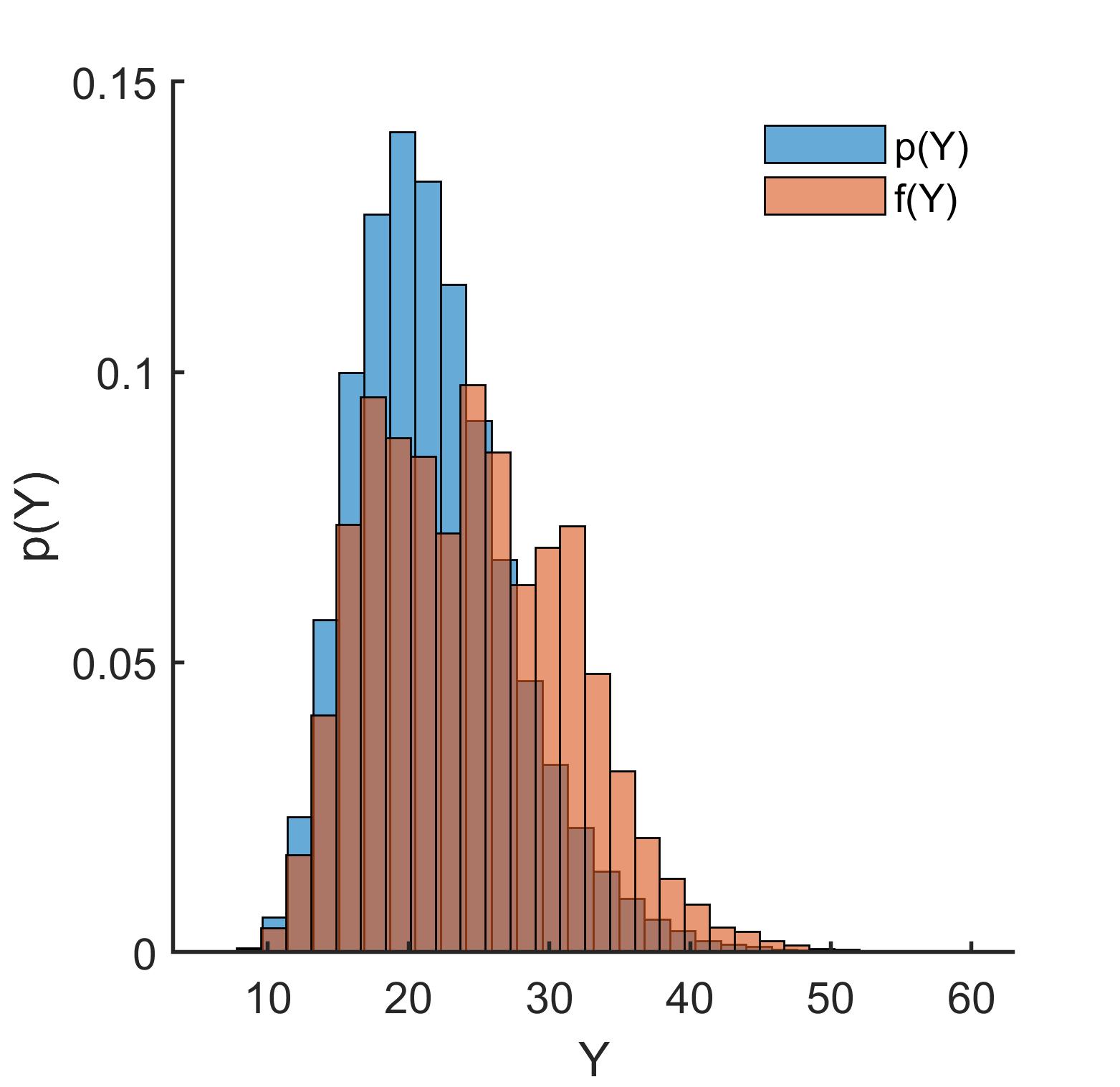} 
\hskip-.25in\includegraphics[width=3.4in]{\figdir/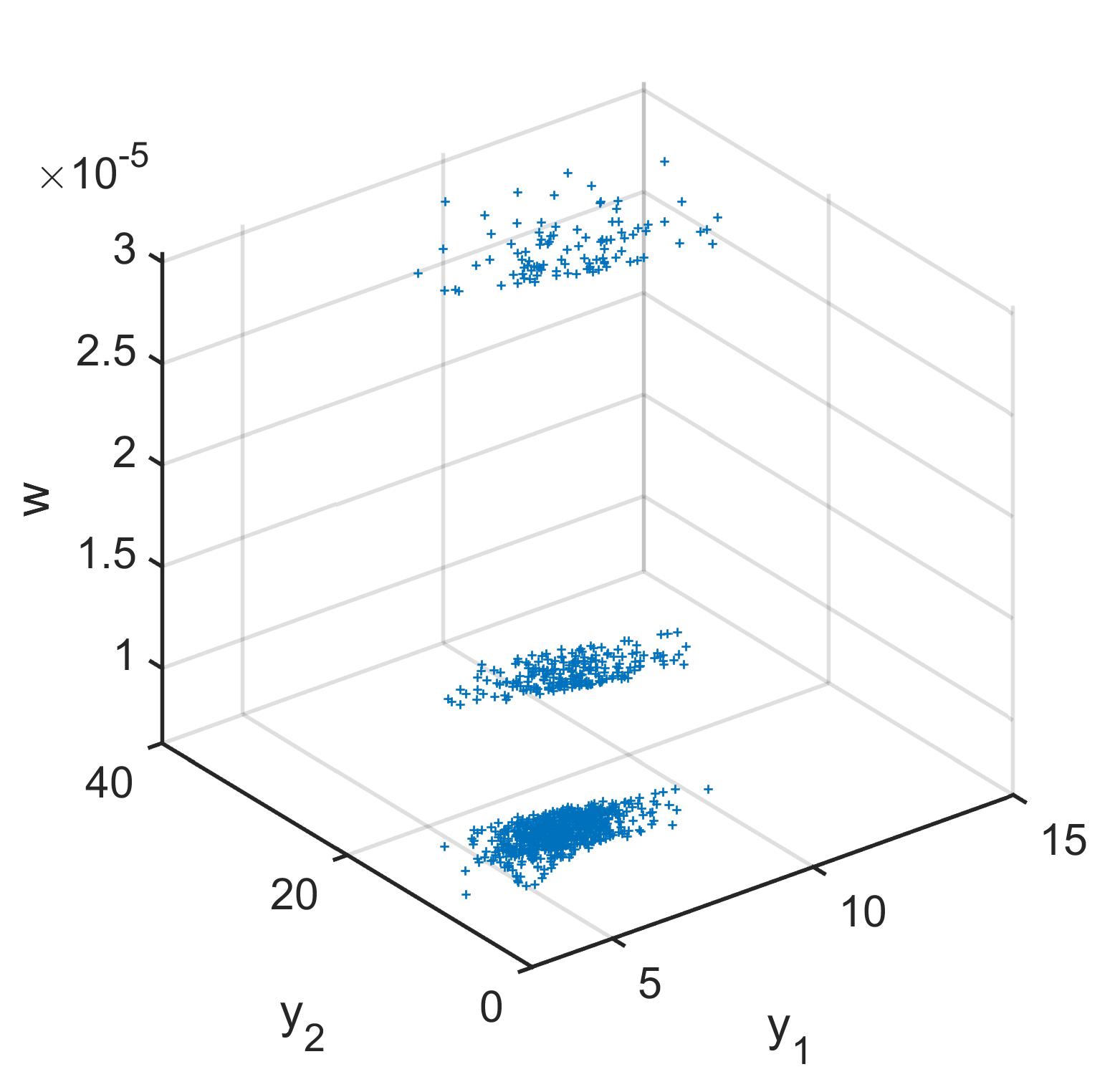} \\
    \caption{In Example 1, $\y$ is in $m=2$ dimensions with a positively dependent bivariate lognormal $p(\y)$ and 
     quantile constraints on the total $Y=\bone'\y$.  Constraints favour somewhat higher values of $Y$ than predicted under $p(\y)$. 
     \\ {\bf\blu Upper row:}  {\em Left:} Contours of $p(\y)$ with a random subsample from the ET constrained $f(\y)$ indicated by IS weights (multiplied by Monte Carlo sample size). 
      {\em Right:} Marginal p.d.f.s $p(y_j)$ overlaid on histograms of reweighted samples representing the $f_j(y_j)$ for $j=\seq 12.$
    \\ {\bf\blu Lower row:} {\em Left:}   Histograms representing $p(Y)$ and $f(Y)$.  {\em Right:}  Scatter plot of a random sample of the IS weights $w_i$ against $y_1$ and $y_2$.   
    \label{fig:TotalYrhoPosA}}
\end{figure}

\parablu{Example 1.}
The example has $\phi(\y)=Y = \bone'\y$   where $\y$ is bivarariate lognormal with positive dependence between $y_1$ and $y_2$. See Figure~\ref{fig:TotalYrhoPosA} for graphical summaries noted here.   Quantile constraints are on the three quartiles of $Y$ set at somewhat higher  values relative to those initially predicted;  specifically, the lower quartile, median and upper quartile of $Y$ are constrained to be 5, 15 and 20\% higher than those under  $p(Y),$ respectively.   

In the quantile constraint setting generally,  the ET weight function $\exp(\btau'\s(\y))$ takes values in a discrete set since $\s(\y)$ is a vector of indicators. 
With $q=3$ this yields four possible values, and the specific selection of constraints on the 3 quartiles is a special case.  With a random sample from $p(\y)$ the resulting IS weights to convert to samples from $f(\y)$  then take values from the set of 4 distinct weights evaluated at the optimizing $\btau.$ 
 Figure~\ref{fig:TotalYrhoPosA} provides summaries.   A large Monte Carlo sample from $p(\y)$ is reweighted with IS weights, with a small (random) subsample of the draws indicated as a scatter plot overlaid on the baseline $p(\y).$ These draws are indexed by the value $W = I w$ where $w$ is the normalized IS weight and I the Monte Carlo sample size.   Since the ET constraints here promote larger values of the sum of $y_1$ and $y_2$, the weights are higher for larger values of each of the elements of $\y.$       The example here has an ESS of around 70\%. 
 
\begin{figure}[htp!]
\centering
\hskip-.15in\includegraphics[width=3.4in]{\figdir/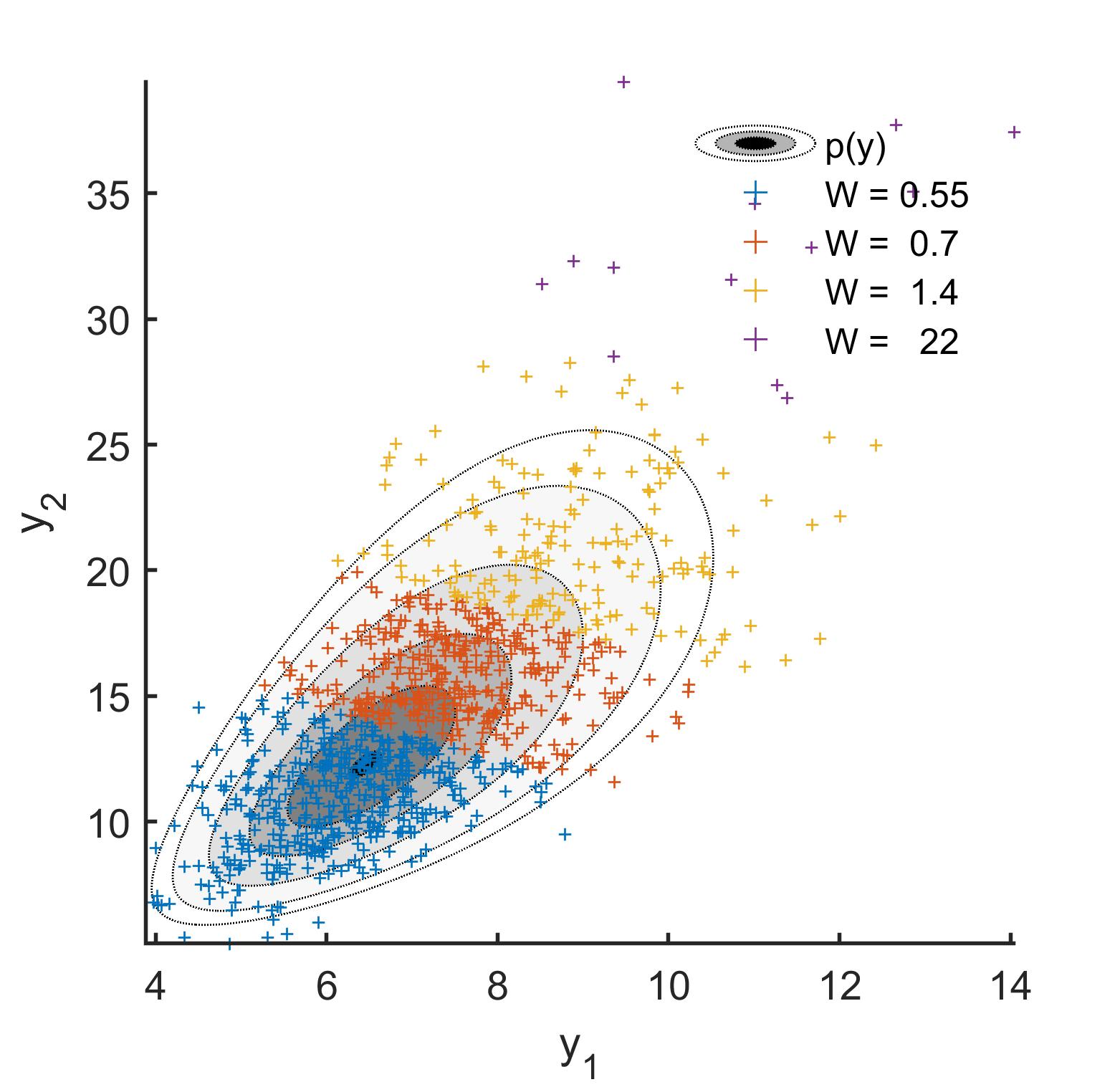} 
\hskip-.25in\includegraphics[width=3.4in]{\figdir/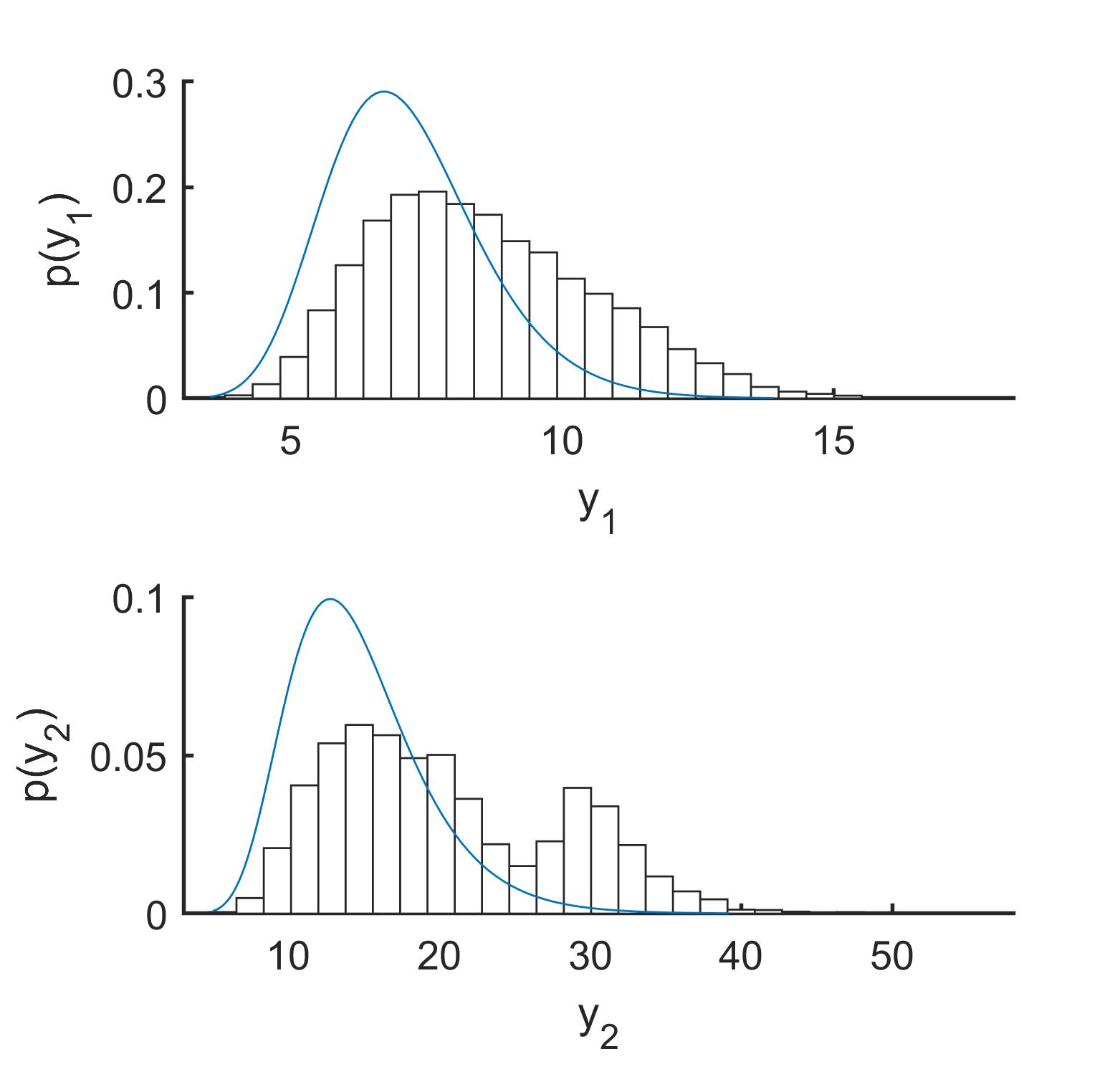} \\
\hskip-.15in\includegraphics[width=3.4in]{\figdir/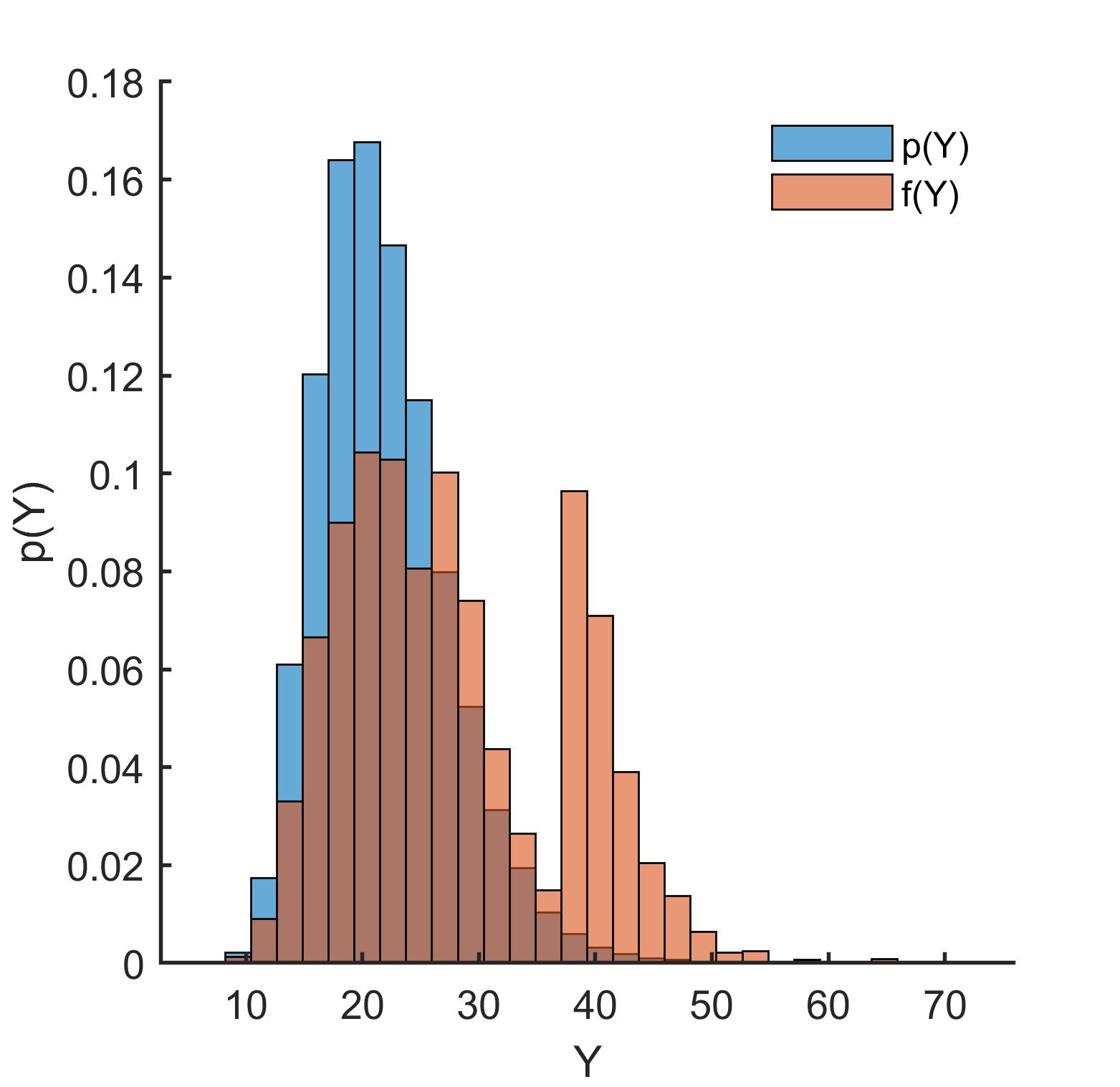} 
\hskip-.25in\includegraphics[width=3.4in]{\figdir/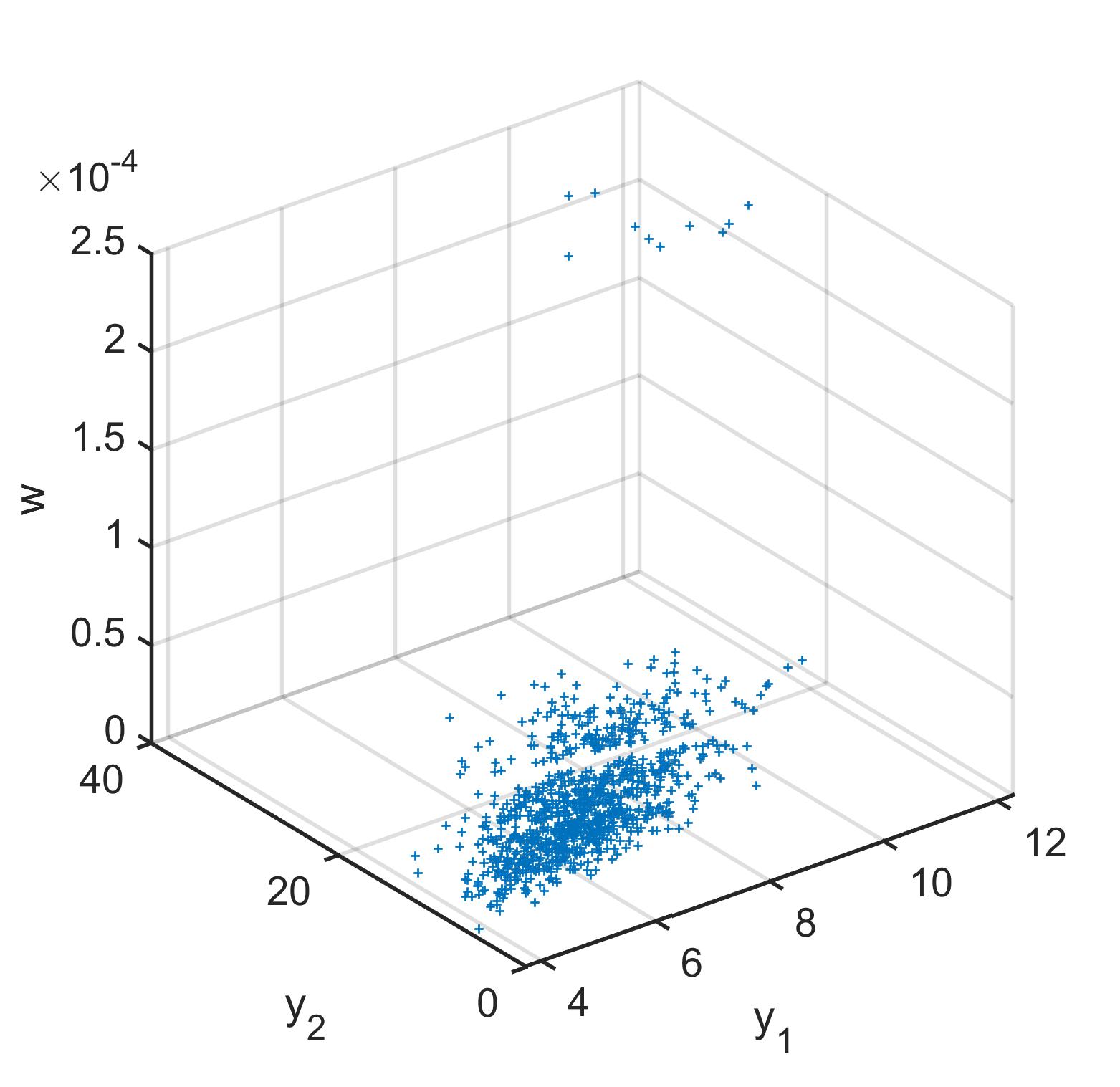} \\
    \caption{In Example 2, the model and analysis are as in Example 1 and Figure~\ref{fig:TotalYrhoPosA} but with quartile constraints on the total $Y=\bone'\y$ now favouring  higher values. 
    \label{fig:TotalYrhoPosB}}
\end{figure}

\parablu{Example 2.} 
This example repeats that of Example 1 but with a more aggressive set of quantile constraints favouring higher values of the total $Y$.   Here 
the lower quartile, median and upper quartile of $Y$ are constrained to be 15, 25 and 50\% higher than those under the baseline    $p(Y).$  See Figure~\ref{fig:TotalYrhoPosB}. All other details are as in Example 1.  Here the ESS is around 17\% showing increased conflict between the imposed constraints and the baseline predictive $p(\y)$ relative to that with the less extreme constraints in Example 1.   

\begin{figure}[htp!]
\centering
\hskip-.15in\includegraphics[width=3.4in]{\figdir/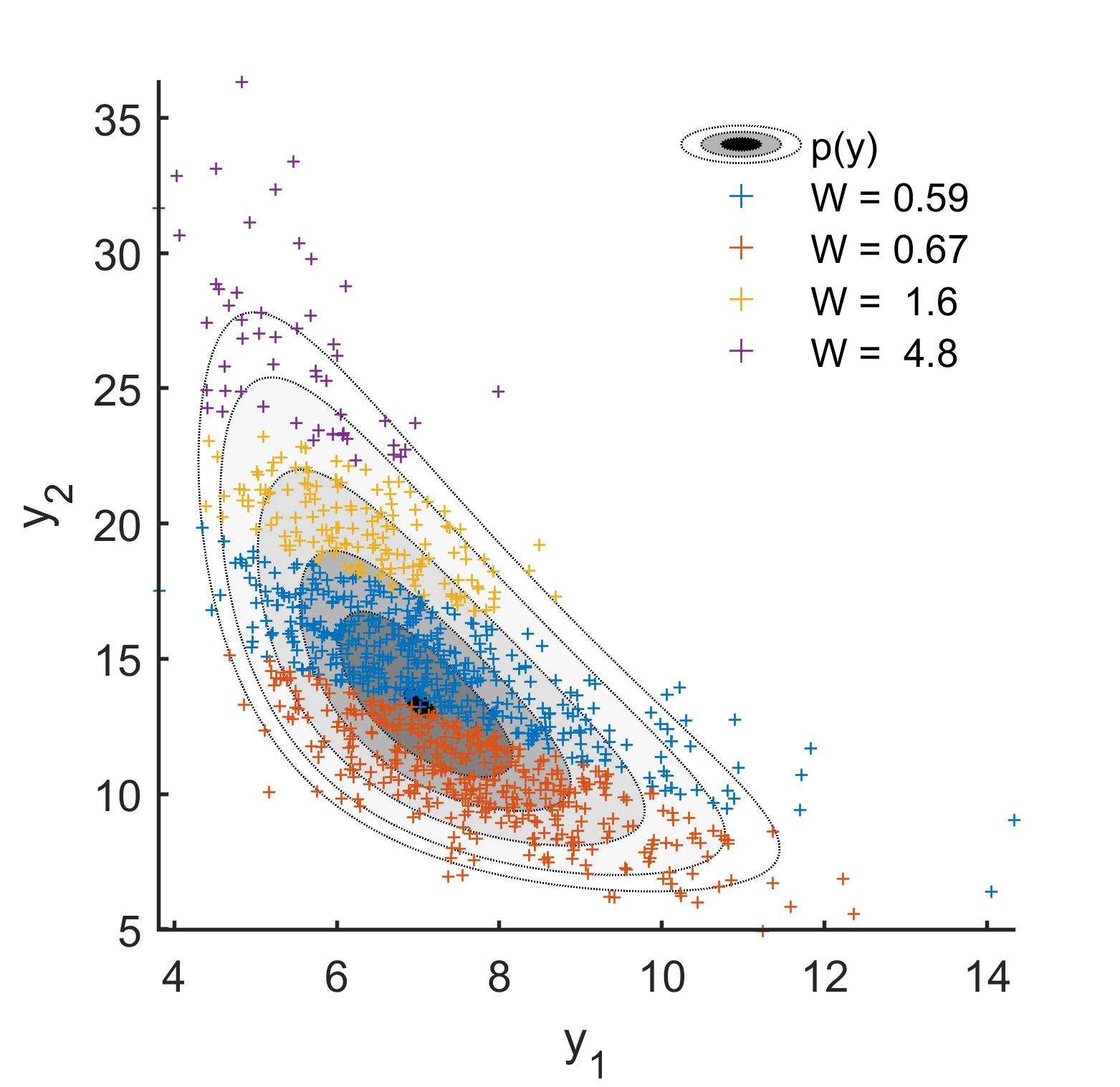} 
\hskip-.25in\includegraphics[width=3.4in]{\figdir/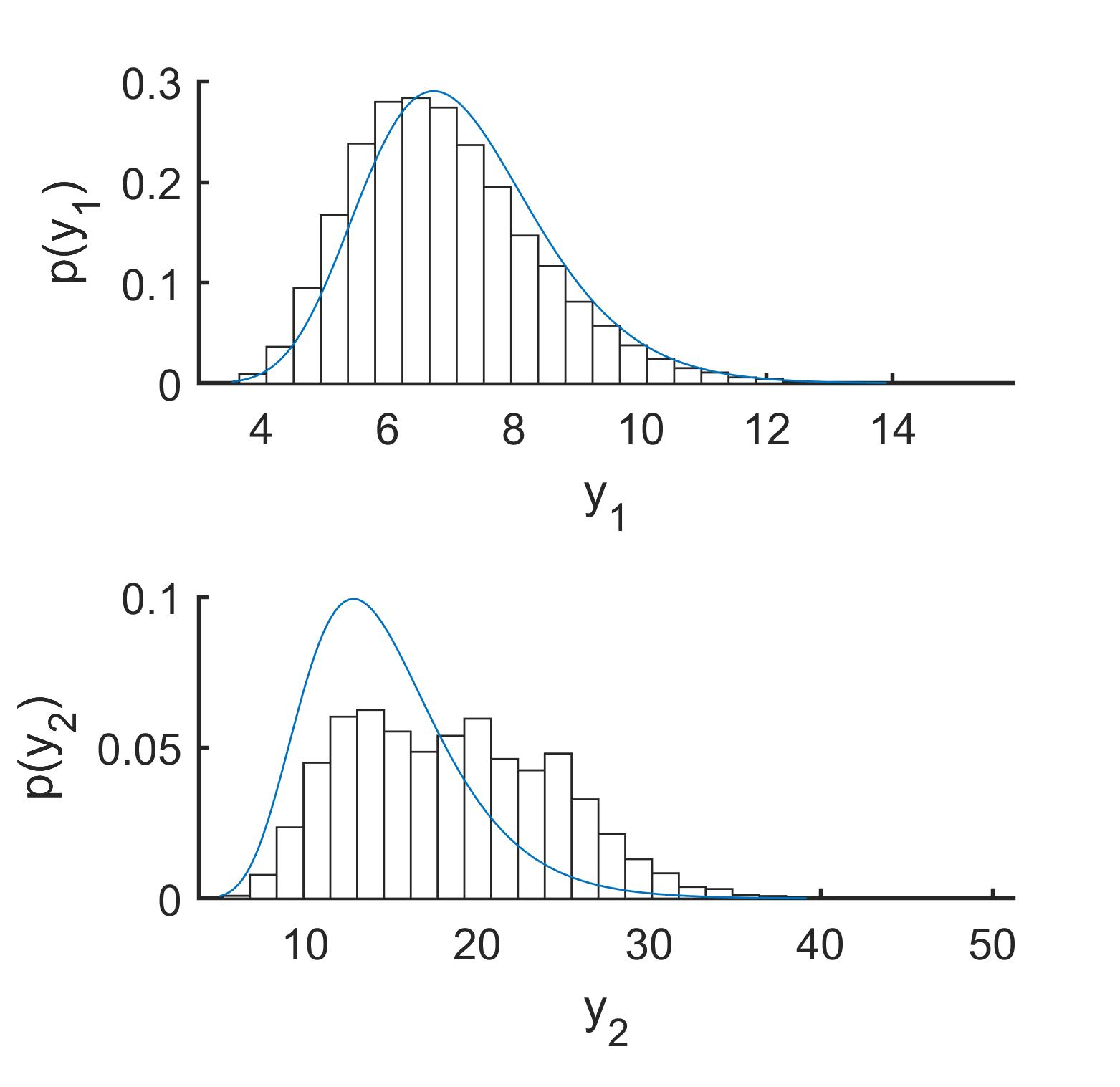} \\
\hskip-.15in\includegraphics[width=3.4in]{\figdir/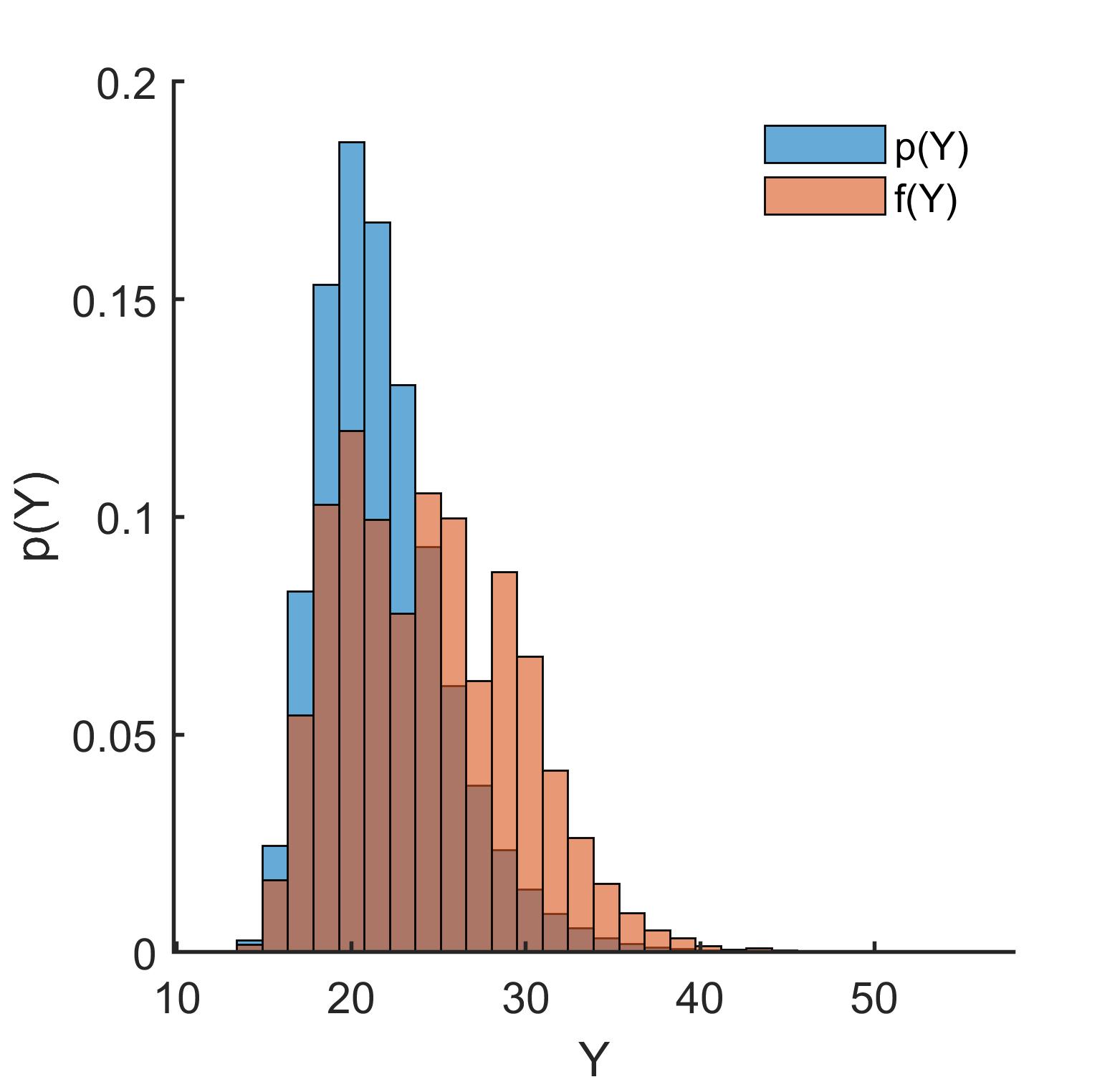} 
\hskip-.25in\includegraphics[width=3.4in]{\figdir/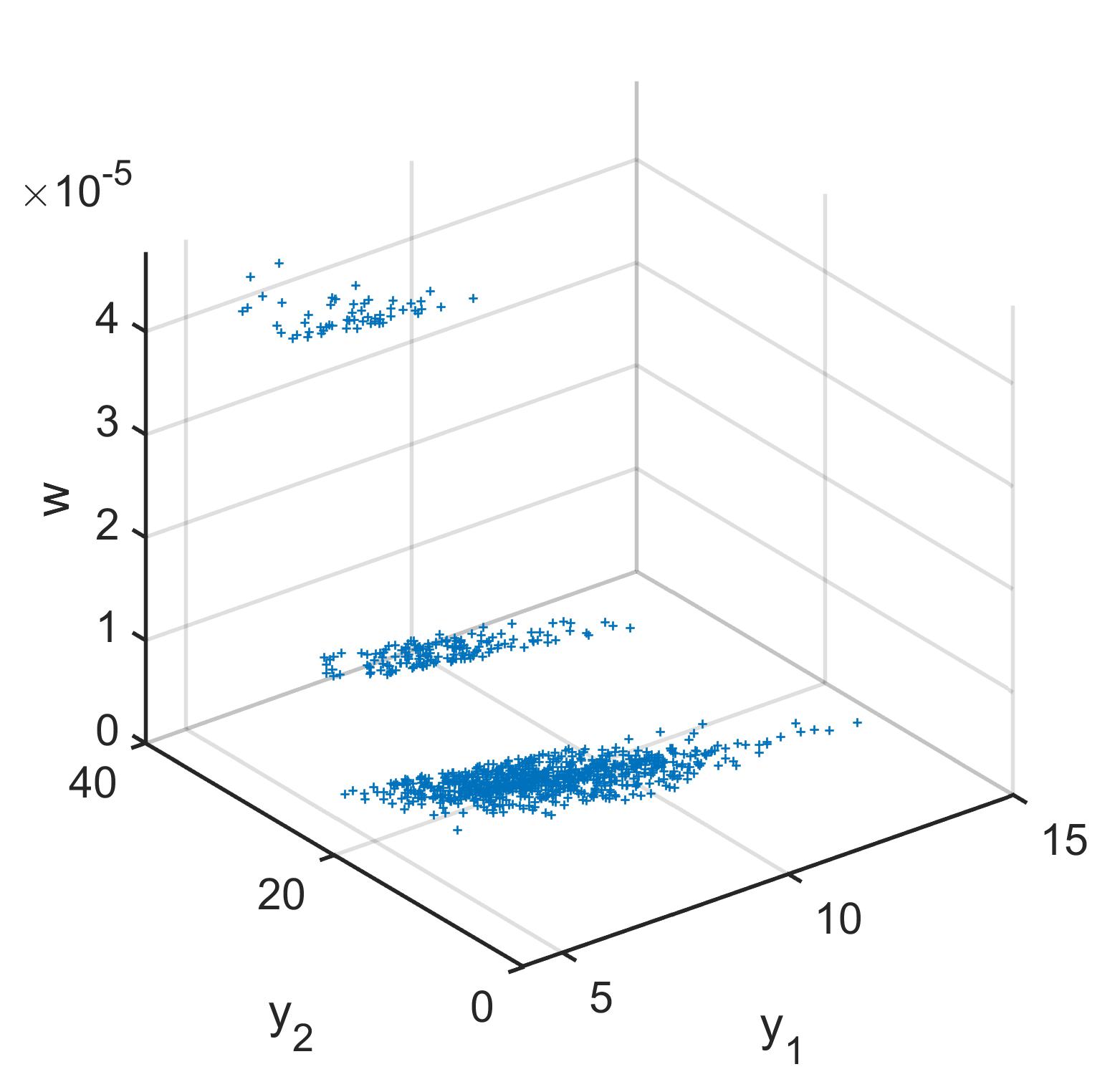} \\
    \caption{In Example 3, the model and analysis are as in Example 1 and Figure~\ref{fig:TotalYrhoPosA} but now with negative dependence between $y_1$ and $y_2$ in the baseline   predictive $p(\y).$
    \label{fig:TotalYrhoNegC}}
\end{figure}

\parablu{Example 3.} 
This example repeats that of Example 1 but with negative dependence in the baseline $(\y).$    Here the ESS is around 53\% showing increased conflict between the imposed constraints and the baseline predictive $p(\y)$ relative to that with positive dependence in Example 1. See Figure~\ref{fig:TotalYrhoNegC}.

\begin{figure}[htp!]
\centering
\hskip-.15in\includegraphics[width=3.4in]{\figdir/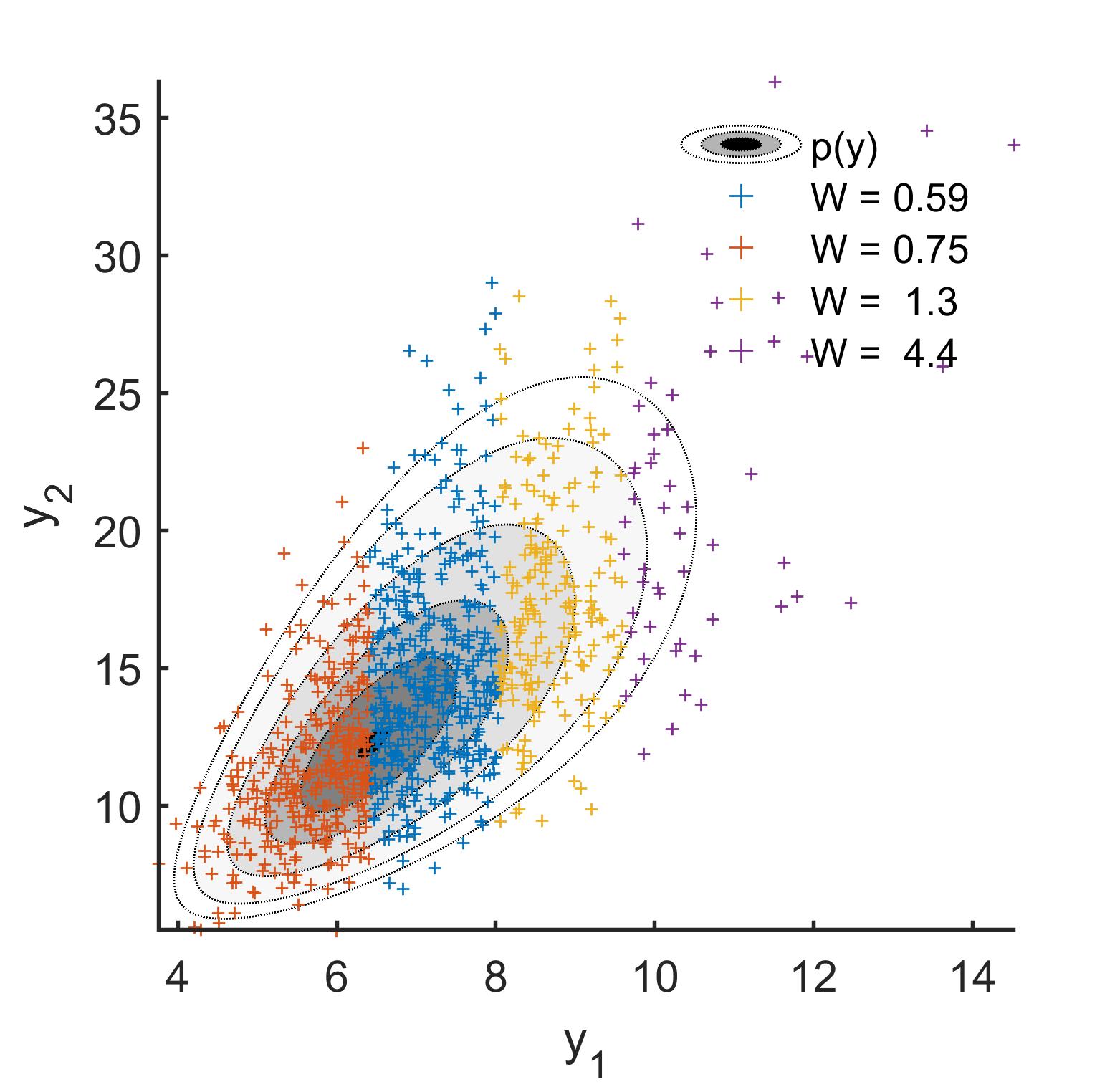} 
\hskip-.25in\includegraphics[width=3.4in]{\figdir/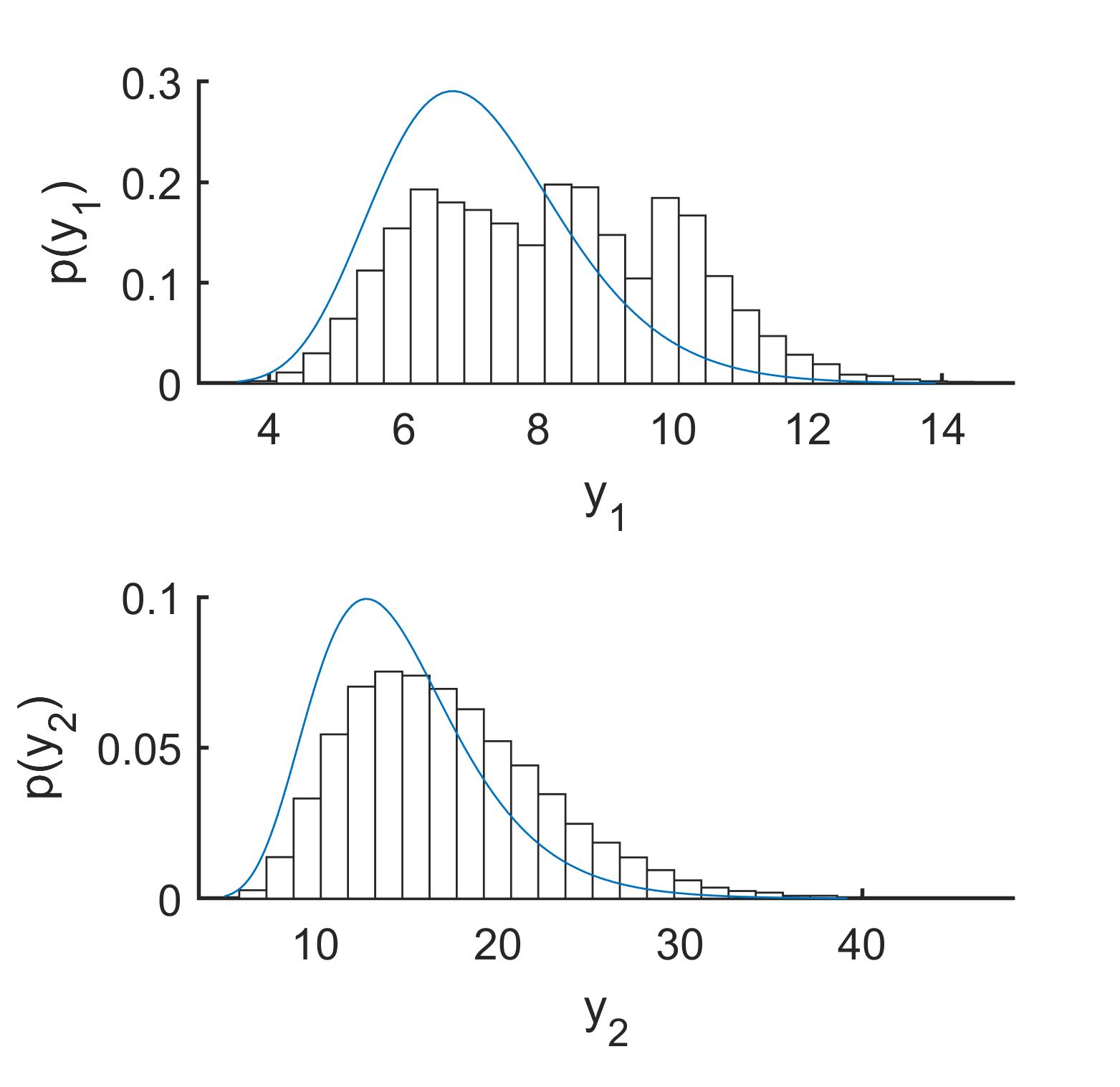} \\
\hskip-.15in\includegraphics[width=3.4in]{\figdir/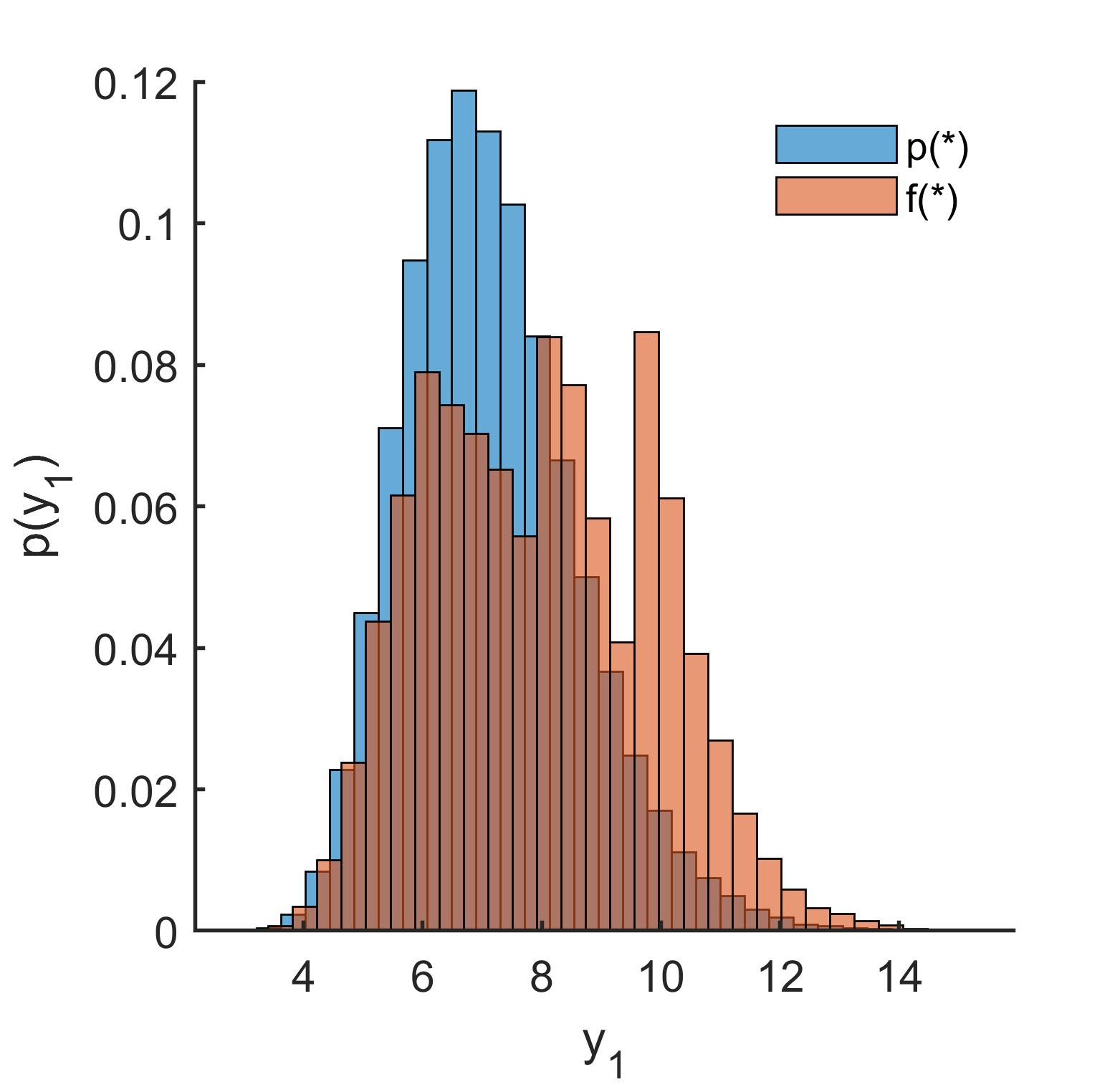} 
\hskip-.25in\includegraphics[width=3.4in]{\figdir/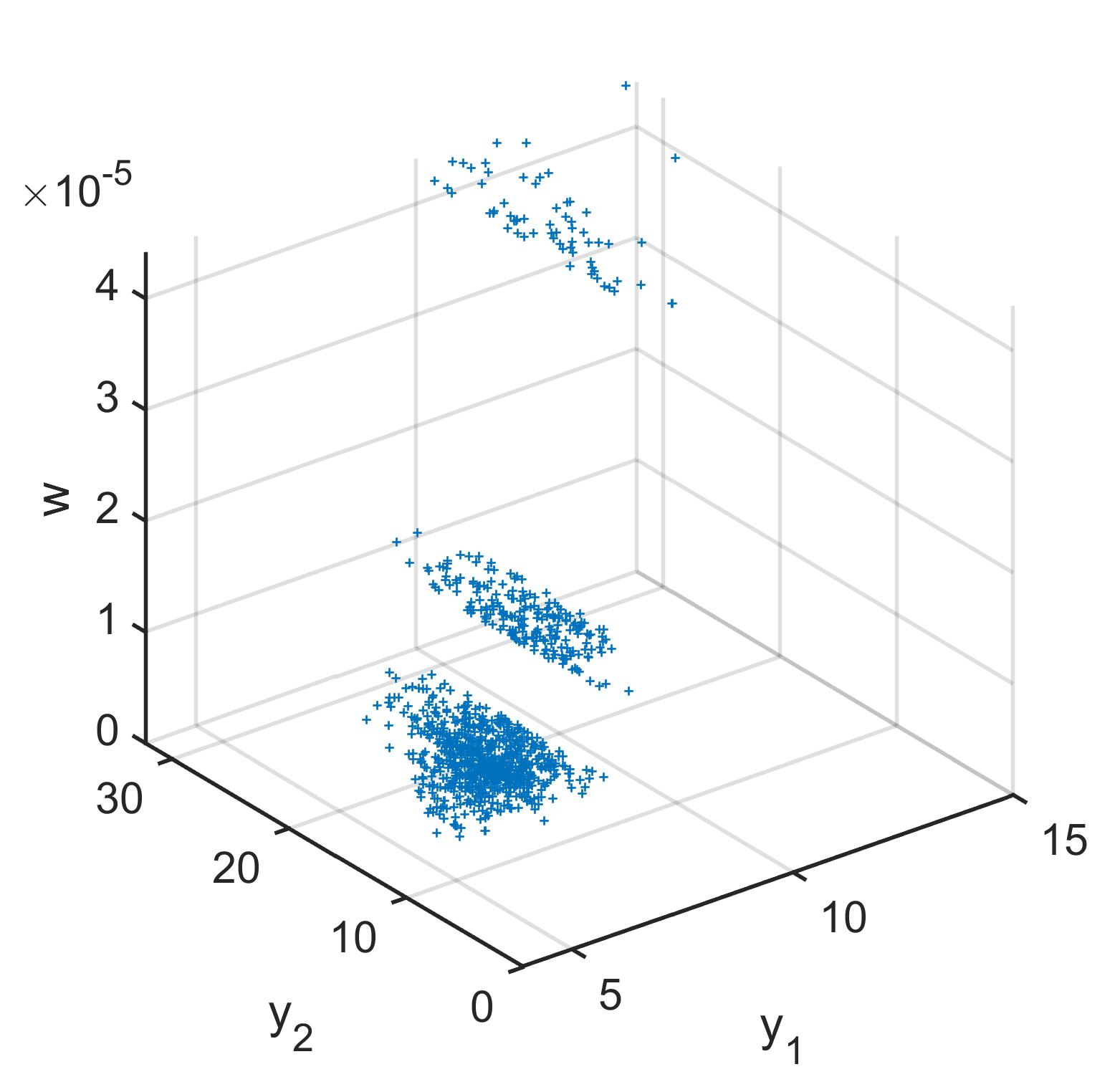} \\
    \caption{In Example 4, the model and analysis are as in Example 1 and Figure~\ref{fig:TotalYrhoPosA} but now with quartile constraints on the marginal distribution of  $y_1$. 
    \label{fig:y1rhoPos}}
\end{figure}

\parablu{Example 4.} 
This example is as in Example 1 but now has $\phi(\y)=y_1$ and so imposes quartile constraints on the ET margin $f_1(y_1)$ only.  The three quartiles of $y_1$  are constrained to be 5, 15 and 20\% higher than those under  $p_1(y_1).$  All other details are as in Example 1. Here the ESS is around 56\%.  See Figure~\ref{fig:y1rhoPos}.

\begin{figure}[htp!]
\centering
\hskip-.15in\includegraphics[width=3.4in]{\figdir/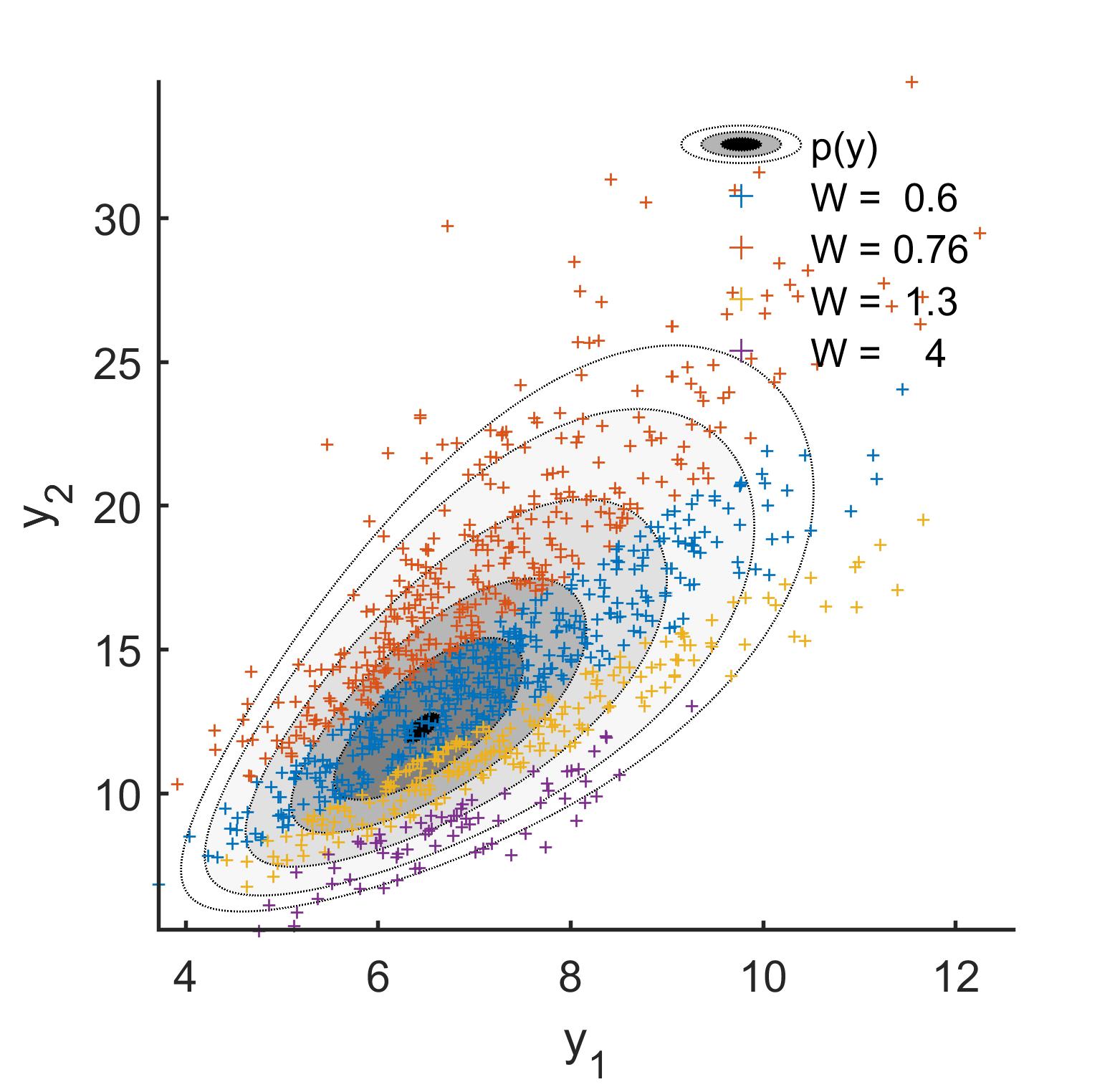} 
\hskip-.25in\includegraphics[width=3.4in]{\figdir/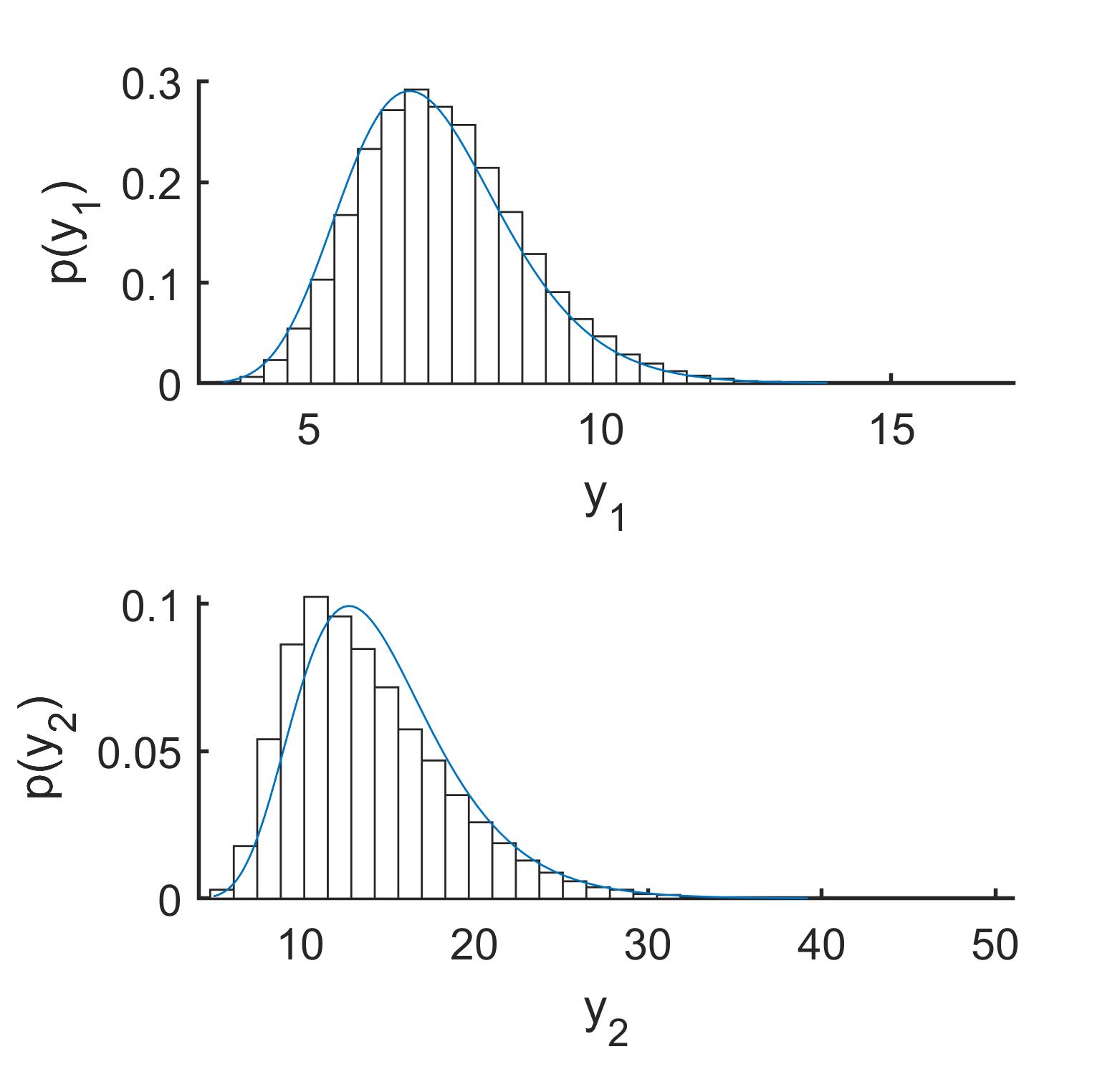} \\
\hskip-.15in\includegraphics[width=3.4in]{\figdir/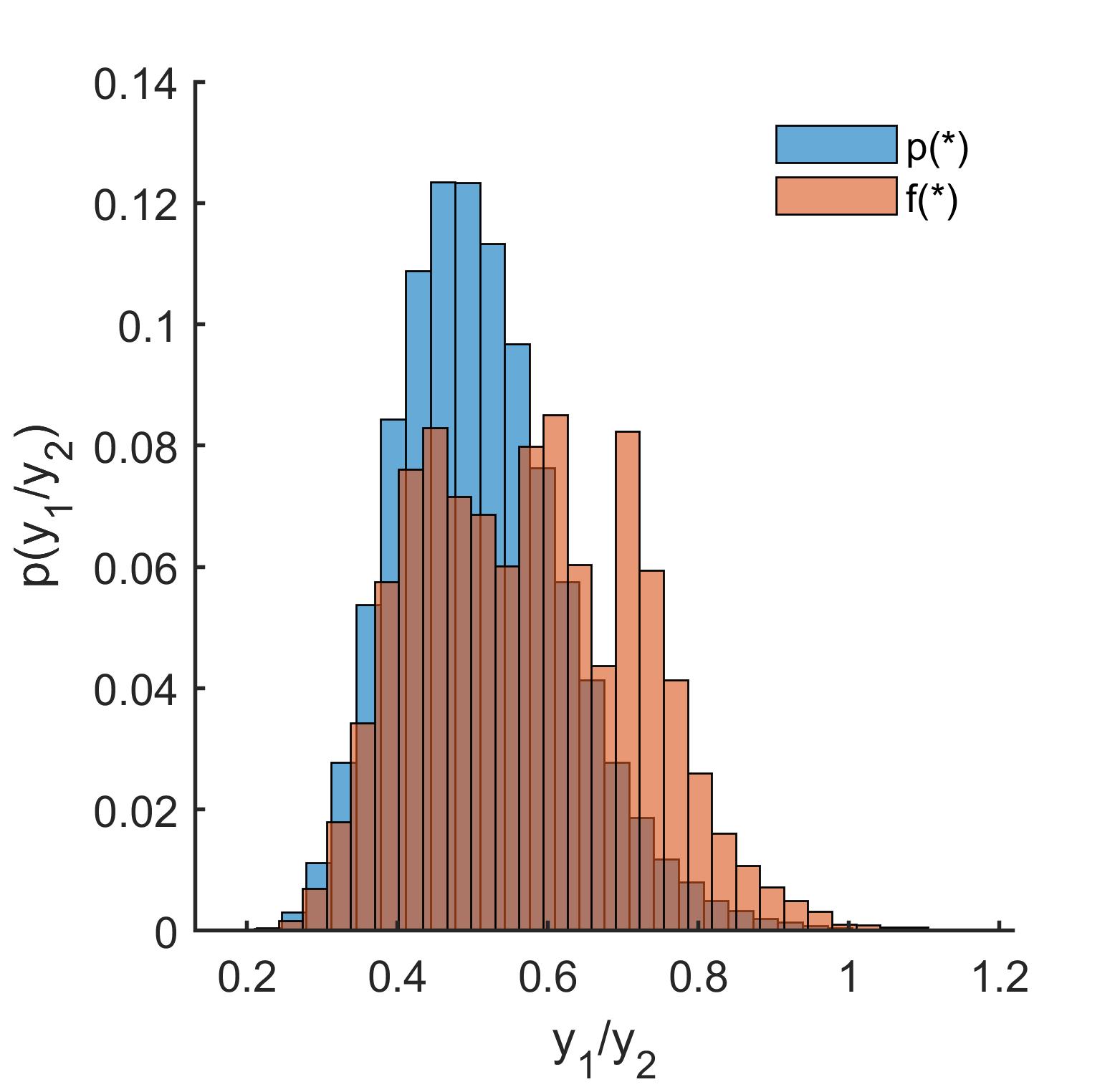} 
\hskip-.25in\includegraphics[width=3.4in]{\figdir/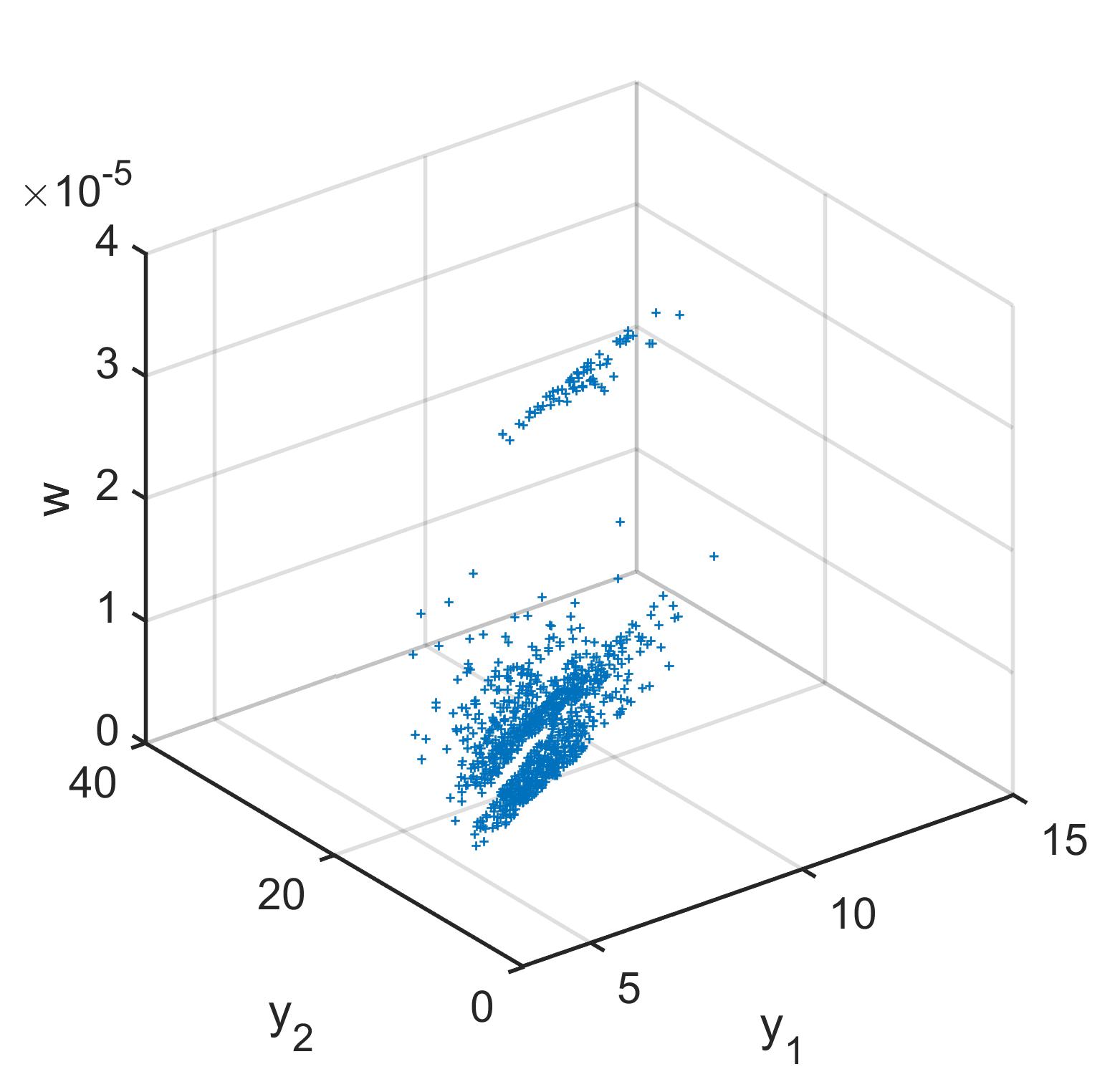} \\
    \caption{In Example 5, the model and analysis are as in Example 1 and Figure~\ref{fig:TotalYrhoPosA} but now with quartile constraints on the marginal distribution of  the ratio $y_1/y_2$. 
    \label{fig:y1-over-y2-rhoPos}}
\end{figure}

\parablu{Example 5.} 
This example is as in Example 1 but now has $\phi(\y)=y_1/y_2$ and so imposes quartile constraints on the ET margin of this relative value.  The three quartiles are constrained to be 5, 15 and 20\% higher than those implied by  $p(\y).$  All other details are as in Example 1. Here the ESS is around 60\%. See Figure~\ref{fig:y1-over-y2-rhoPos}.

\FloatBarrier

\section*{Acknowledgements}
This research was partially supported by $84.51^\circ$ Labs.  The authors acknowledge Christoph Hellmayr $(84.51^\circ$~Labs.),  Joseph Lawson and Graham Tierney (Department of Statistical Science, Duke University) for thoughtful discussions that helped with our perspectives on material developed and presented here. Earlier discussions with Domenico Giannoni (Amazon) helped to stimulate our interests in the research areas contacted here. 

\renewcommand{\refname}{\large\blu References}
\small
\bibliographystyle{chicago} 
\bibliography{ET}   

\begin{thebibliography}{}

\bibitem[\protect\citeauthoryear{Barndorff-Nielsen}{Barndorff-Nielsen}{1978}]{BarndorffNielsen1978}
Barndorff-Nielsen, O. (1978).
\newblock {\em Information and Exponential Families}.
\newblock Wiley.

\bibitem[\protect\citeauthoryear{Fox}{Fox}{1987}]{FoxFunctionalCalculus1987}
Fox, C. (1987).
\newblock {\em An Introduction to the Calculus of Variations}.
\newblock Courier Dover Publications.

\bibitem[\protect\citeauthoryear{Gruber and West}{Gruber and
  West}{2016}]{GruberWest2016BA}
Gruber, L.~F. and M.~West (2016).
\newblock {GPU}-accelerated {B}ayesian learning and forecasting in simultaneous
  graphical dynamic linear models.
\newblock {\em Bayesian Analysis\/}~{\em 11}, 125--149.

\bibitem[\protect\citeauthoryear{Gruber and West}{Gruber and
  West}{2017}]{GruberWest2017ECOSTA}
Gruber, L.~F. and M.~West (2017).
\newblock Bayesian forecasting and scalable multivariate volatility analysis
  using simultaneous graphical dynamic linear models.
\newblock {\em Econometrics and Statistics\/}~{\em 3}, 3--22.

\bibitem[\protect\citeauthoryear{Jaynes}{Jaynes}{1957}]{Jaynes1957}
Jaynes, E.~T. (1957).
\newblock Information theory and statistical mechanics.
\newblock {\em Physical Review\/}~{\em 106}, 620--630.
\newblock Available at:
  \href{http://bayes.wustl.edu/etj/articles/theory.1.pdf}{http://bayes.wustl.edu/etj/articles/theory.1.pdf}.

\bibitem[\protect\citeauthoryear{Krüger, Clark, and Ravazzolo}{Krüger
  et~al.}{2017}]{KrugerET2017}
Krüger, F., T.~E. Clark, and F.~Ravazzolo (2017).
\newblock Using entropic tilting to combine {BVAR} forecasts with external
  nowcasts.
\newblock {\em Journal of Business and Economic Statistics\/}~{\em 35\/}(3),
  470--485.

\bibitem[\protect\citeauthoryear{Robertson, Tallman, and Whitemanng}{Robertson
  et~al.}{2005}]{RobertsonET2005}
Robertson, J.~C., E.~W. Tallman, and C.~H. Whitemanng (2005).
\newblock Forecasting using relative entropy.
\newblock {\em Journal of Money, Credit, and Banking\/}~{\em 37}, 383--401.

\bibitem[\protect\citeauthoryear{Tallman and West}{Tallman and
  West}{2022}]{TallmanWest2022}
Tallman, E. and M.~West (2022).
\newblock Bayesian predictive decision synthesis.
\newblock {\em Submitted for publication\/}.
\newblock arXiv:2206.03815.

\bibitem[\protect\citeauthoryear{West}{West}{2021}]{West2021decisionconstraints}
West, M. (2021).
\newblock Perspectives on constrained forecasting.
\newblock {\em Invited revision under review\/}.
\newblock arXiv:2007.11037.

\end{thebibliography}
 
  
  \end{document}